\documentstyle[12pt,fleqn,epsfig]{article}

\begin{document}

\begin{center}
  {\LARGE \bf Energy dependence of transverse momentum fluctuations in 
   Pb+Pb collisions at the CERN Super Proton Synchrotron (SPS) at 20$A$ 
   to 158$A$ GeV}

\end{center}

\vspace{0.5cm}

\begin{center}
  {\Large \bf The NA49 Collaboration}
\end{center}

\vspace{0.5cm}

\begin{abstract}

Results are presented on event-by-event fluctuations of transverse 
momenta $p_T$ in central Pb+Pb interactions at 20$A$, 30$A$, 40$A$, 
80$A$, and 158$A$ GeV. The analysis was performed for charged particles 
at forward center-of-mass rapidity ($1.1 < y^{*}_{\pi} < 2.6$). Three 
fluctuation measures were studied: the distribution of average 
transverse momentum - $M(p_T)$ - in the event, the $\Phi_{p_{T}}$ 
fluctuation measure and two-particle transverse momentum correlations. 
Fluctuations of $p_T$ are small and show no significant energy 
dependence in the energy range of the CERN Super Proton Synchrotron.  
Results are compared with QCD-inspired predictions for the critical 
point, and with the UrQMD model. Transverse momentum fluctuations, 
similar to multiplicity fluctuations, do not show the increase expected 
for freeze-out near the critical point of QCD.

\end{abstract}

\newpage

\begin{center}
  {\bf The NA49 Collaboration}
\end{center}
\vspace{0.5cm}
\noindent
T.~Anticic$^{23}$, B.~Baatar$^{8}$,D.~Barna$^{4}$,
J.~Bartke$^{6}$, L.~Betev$^{10}$, H.~Bia{\l}\-kowska$^{20}$,
C.~Blume$^{9}$,  B.~Boimska$^{20}$, M.~Botje$^{1}$,
J.~Bracinik$^{3}$, P.~Bun\v{c}i\'{c}$^{10}$,
V.~Cerny$^{3}$, P.~Christakoglou$^{2}$,
P.~Chung$^{19}$, O.~Chvala$^{14}$,
J.G.~Cramer$^{16}$, P.~Csat\'{o}$^{4}$, P.~Dinkelaker$^{9}$,
V.~Eckardt$^{13}$,
Z.~Fodor$^{4}$, P.~Foka$^{7}$,
V.~Friese$^{7}$, J.~G\'{a}l$^{4}$,
M.~Ga\'zdzicki$^{9,11}$, V.~Genchev$^{18}$, 
E.~G{\l}adysz$^{6}$, K.~Grebieszkow$^{22}$,
S.~Hegyi$^{4}$, C.~H\"{o}hne$^{7}$,
K.~Kadija$^{23}$, A.~Karev$^{13}$, D.~Kikola$^{22}$,
V.I.~Kolesnikov$^{8}$, E.~Kornas$^{6}$,
R.~Korus$^{11}$, M.~Kowalski$^{6}$,
M.~Kreps$^{3}$, A.~Laszlo$^{4}$,
R.~Lacey$^{19}$, M.~van~Leeuwen$^{1}$,
P.~L\'{e}vai$^{4}$, L.~Litov$^{17}$, B.~Lungwitz$^{9}$,
M.~Makariev$^{17}$, A.I.~Malakhov$^{8}$,
M.~Mateev$^{17}$, G.L.~Melkumov$^{8}$, A.~Mischke$^{1}$, 
M.~Mitrovski$^{9}$,
J.~Moln\'{a}r$^{4}$, St.~Mr\'owczy\'nski$^{11}$, V.~Nicolic$^{23}$,
G.~P\'{a}lla$^{4}$, A.D.~Panagiotou$^{2}$, D.~Panayotov$^{17}$,
A.~Petridis$^{2,\ast}$, W.~Peryt$^{22}$, M.~Pikna$^{3}$, 
J.~Pluta$^{22}$,
D.~Prindle$^{16}$,
F.~P\"{u}hlhofer$^{12}$, R.~Renfordt$^{9}$,
C.~Roland$^{5}$, G.~Roland$^{5}$,
M. Rybczy\'nski$^{11}$, A.~Rybicki$^{6}$,
A.~Sandoval$^{7}$, N.~Schmitz$^{13}$, T.~Schuster$^{9}$, 
P.~Seyboth$^{13}$,
F.~Sikl\'{e}r$^{4}$, B.~Sitar$^{3}$, E.~Skrzypczak$^{21}$, 
M.~Slodkowski$^{22}$,
G.~Stefanek$^{11}$, R.~Stock$^{9}$, C.~Strabel$^{9}$, 
H.~Str\"{o}bele$^{9}$, T.~Susa$^{23}$,
I.~Szentp\'{e}tery$^{4}$, J.~Sziklai$^{4}$, M.~Szuba$^{22}$, 
P.~Szymanski$^{10,20}$,
V.~Trubnikov$^{20}$, M.~Utvic$^{9}$, D.~Varga$^{4,10}$, 
M.~Vassiliou$^{2}$, G.I.~Veres$^{4,5}$, G.~Vesztergombi$^{4}$,
D.~Vrani\'{c}$^{7}$, 
Z.~W{\l}odarczyk$^{11}$, A.~Wojtaszek$^{11}$, I.K.~Yoo$^{15}$

\vspace{0.5cm}
\noindent
$^{1}$NIKHEF, Amsterdam, Netherlands. \\
$^{2}$Department of Physics, University of Athens, Athens, Greece.\\
$^{3}$Comenius University, Bratislava, Slovakia.\\
$^{4}$KFKI Research Institute for Particle and Nuclear Physics, 
Budapest, Hungary.\\
$^{5}$MIT, Cambridge, USA.\\
$^{6}$Institute of Nuclear Physics, Cracow, Poland.\\
$^{7}$Gesellschaft f\"{u}r Schwerionenforschung (GSI), Darmstadt, 
Germany.\\
$^{8}$Joint Institute for Nuclear Research, Dubna, Russia.\\
$^{9}$Fachbereich Physik der Universit\"{a}t, Frankfurt, Germany.\\
$^{10}$CERN, Geneva, Switzerland.\\
$^{11}$Institute of Physics, Jan Kochanowski University, Kielce, 
Poland.\\
$^{12}$Fachbereich Physik der Universit\"{a}t, Marburg, Germany.\\
$^{13}$Max-Planck-Institut f\"{u}r Physik, Munich, Germany.\\
$^{14}$Institute of Particle and Nuclear Physics, Charles University, 
Prague, Czech Republic.\\
$^{15}$Department of Physics, Pusan National University, Pusan, Republic 
of Korea.\\
$^{16}$Nuclear Physics Laboratory, University of Washington, Seattle, 
WA, USA.\\
$^{17}$Atomic Physics Department, Sofia University St. Kliment Ohridski, 
Sofia, Bulgaria.\\
$^{18}$Institute for Nuclear Research and Nuclear Energy, Sofia, 
Bulgaria.\\
$^{19}$Department of Chemistry, Stony Brook Univ. (SUNYSB), Stony Brook, 
USA.\\
$^{20}$Institute for Nuclear Studies, Warsaw, Poland.\\
$^{21}$Institute for Experimental Physics, University of Warsaw, Warsaw, 
Poland.\\
$^{22}$Faculty of Physics, Warsaw University of Technology, Warsaw, 
Poland.\\
$^{23}$Rudjer Boskovic Institute, Zagreb, Croatia.\\
$^{\ast}$deceased

\newpage

\section{Introduction and Motivation}

\indent
For more than 30 years, experiments studying relativistic 
nucleus-nucleus ($A+A$) collisions have been carried out in laboratories 
in Europe and the United States. The main motivation has been to test the 
hypothesis that strongly interacting matter at energy densities 
exceeding about 1 GeV/fm$^{3}$ exists in the form of deconfined quarks 
and gluons, eventually forming the quark-gluon plasma 
(QGP) \cite{Col75}. Recent results from the CERN Super Proton 
Synchrotron (SPS) and BNL Relativistic Heavy Ion Collider (RHIC) seem 
to confirm this conjecture. The data suggest that the threshold for the 
onset of deconfinement is located at the low SPS energies 
\cite{mg_model, na49_kpi}.

\indent
The phase diagram of strongly interacting matter can be
presented in terms of temperature $T$ and baryochemical
potential $\mu_B$. QCD-inspired calculations
suggest that the phase boundary between hadrons and QGP is
of first order at large values of $\mu_B$, ending in a critical
point of second order and then turning into a continuous
rapid transition at low $\mu_B$ \cite{fodor_latt_2004}. 
The location of the critical point may be close to the $(T, \mu_B)$ 
values found for the freeze-out of the hadron system produced in 
collisions of heavy nuclei at SPS energies.

\indent
The study of fluctuations is an important tool for localizing the phase 
boundary and the critical point. In particular, significant
transverse momentum and multiplicity fluctuations are expected
to appear for systems that hadronize from QGP and freeze-out
near the critical point of QCD \cite{SRS}. The location of the
freeze-out point in the phase diagram can be moved by varying
the collision energy and the size of the collision system. A 
nonmonotonic evolution of fluctuations with these collision
parameters can serve as a signature for the phase transition
and the critical point. These considerations motivated an
extensive program of fluctuation studies at the SPS and RHIC
accelerators.

\indent
So far for central Pb+Pb collisions, the analysis of 
multiplicity fluctuations has found only small effects without 
significant structure in the energy dependence in the whole SPS energy
domain \cite{benjamin_2007}. The measured net charge fluctuations
can be explained by the effects of global charge conservation
\cite{jacek_charge}. The energy dependence of event-by-event
fluctuations of the $K/\pi$ ratio, on the other hand, shows an
interesting increase toward lower energies \cite{qm2004_chr} that can
be attributed to the onset of deconfinement \cite{fluct_SMES} rather
than to the critical point.

\indent
This paper presents results of the NA49 experiment from a study of 
transverse momentum $p_T$ fluctuations in central Pb+Pb
collisions at 20$A$, 30$A$, 40$A$, 80$A$, and 158$A$ GeV. It
extends a previous study \cite{fluct_size} that investigated the 
system size dependence of $p_T$ fluctuations at the top SPS energy.

\indent
Fluctuations in nucleus-nucleus collisions are susceptible to
two trivial sources: the finite and fluctuating number of produced
particles and event-by-event fluctuations of the collision geometry (see 
the discussion of this point within HSD and UrQMD transport
models in Ref. \cite{transport_fluct}). Suitable statistical
tools have to be chosen to extract the fluctuations of interest.
As in the previous NA49 study \cite{fluct_size}, mainly the global 
$\Phi_{p_T}$ measure will be used. Alternative measures used in the 
literature, e.g. $\sigma_{p_{T},dyn}$ \cite{sigmaVolosh}, $\Delta 
\sigma_{p_{T}}$ \cite{Tra00}, $F_{p_T}$ \cite{Fdef}, and 
$\Sigma_{p_{T}} (\%)$ \cite{star2002, CERES} are related \cite{gavin1}.

\indent
To obtain further information on the possible source of 
fluctuations, two additional methods will
be employed. The distribution of event-wise average transverse
momentum $M(p_T)$ will be compared against the corresponding histogram 
for mixed events, which represents purely statistical fluctuations.
Moreover, two-particle transverse momentum correlations, as
proposed in Ref. \cite{Tra00}, will be analyzed.

\indent
This paper is organized as follows. In Sec.~\ref{s:measures} the 
statistical tools used in this analysis are introduced and briefly 
discussed. The NA49 setup is presented in Sec.~\ref{s:experiment}. 
Experimental effects such as detector acceptance and two-track
resolution are discussed in Sec.~\ref{s:data}. The NA49 results on the 
energy dependence of transverse momentum fluctuations are presented and 
discussed in Sec.~\ref{s:results}. A summary closes the paper.

\section{Measures of fluctuations}
\label{s:measures}

\indent
Various methods can be used to measure event-by-event $p_T$ 
fluctuations. A natural observable is the distribution of the average 
transverse momentum of the events defined as

\begin{equation}
M(p_{T})=\frac{1}{N}\sum_{i=1}^{N}p_{Ti},
\end{equation}
where $N$ is the multiplicity of accepted particles in a given 
event and $p_{Ti}$ is the transverse momentum of the $i$-th particle.
The distribution of $M(p_T)$ will be compared with
the corresponding histogram obtained for artificially produced "mixed 
events". In mixed events, all particles are by construction uncorrelated
(each particle in a given mixed event is taken from a different real 
event with the same multiplicity) but follow the experimental inclusive 
three-momentum spectra as well as the distribution of multiplicity.

\indent
The second observable used in this work is the $\Phi_{p_{T}}$ measure, 
proposed in Ref. \cite{Gaz92} and used also in our previous analysis 
\cite{fluct_size}. Following the authors of Ref. \cite{Gaz92}, one can
define the single-particle variable $z_{p_{T}}=p_{T}-\overline{p_{T}}$
with the bar denoting averaging over the single-particle inclusive
distribution. As seen, $\overline{z_{p_{T}}} = 0$.
Further, one introduces the event variable $Z_{p_{T}}$, which is a
multiparticle analog of $z_{p_{T}}$, defined as
\begin{equation}
Z_{p_{T}}=\sum_{i=1}^{N}(p_{Ti}-\overline{p_{T}}),
\end{equation}
where the summation runs over particles in a given event. Note, that   
$\langle Z_{p_{T}} \rangle = 0$, where $\langle ... \rangle$ represents
averaging over events. Finally, the $\Phi_{p_{T}}$ measure is defined
as

\begin{equation}
\label{Phi}
\Phi_{p_{T}}=\sqrt{\frac{\langle
Z_{p_{T}}^{2} \rangle }{\langle N
\rangle }}-\sqrt{\overline{z_{p_{T}}^{2}}}.
\label{eq_phi}
\end{equation}

\indent
The $\Phi_{p_{T}}$ measure has two important properties. 
First, $\Phi_{p_{T}}$ vanishes when  
the system consists of particles that are emitted independently (no 
interparticle correlations), and the single particle momentum spectrum is 
independent of multiplicity. Second, if an $A+A$ collision can be 
treated as an incoherent superposition of independent $N+N$ interactions 
(superposition model), then $\Phi_{p_{T}}$ has a constant value, the same 
for $A+A$ and $N+N$ interactions. This implies that, in particular, 
$\Phi_{p_{T}}$ does not depend on the impact parameter (centrality), if 
the $A+A$ collision is a simple superposition of $N+N$ interactions. 
Furthermore, $\Phi_{p_{T}}$ is independent of
changes of the size of a compact acceptance domain provided 
the correlation scale (range), $l_C$, is much smaller than the size
of the acceptance region, $l_A$. In the limit of large correlation
length and small acceptance the magnitude of $\Phi_{p_T}$ is 
proportional to the multiplicity of accepted particles. Thus for a 
small acceptance $\Phi_{p_T}$ approaches zero. The approximately linear 
dependence of $\Phi_{p_{T}}$ on the fraction of accepted particles 
in the limit $l_A \ll l_C$ suggested the introduction of 
the fluctuation measures $\Sigma_{p_{T}} (\%)$ \cite{CERES} and 
$\sigma_{p_{T},dyn}$ \cite{sigmaVolosh}, which for large particle 
multiplicities are proportional to $\sqrt{\Phi_{p_{T}} / \langle N 
\rangle }$. These properties of the fluctuation measures should be 
taken into account when results are compared and discussed. 
Finally, we note that the magnitude of $\Phi_{p_T}$ decreases 
with the fraction of randomly lost particles, e.g., due to 
incomplete reconstruction efficiency.

\indent
In spite of the above mentioned advantages, there is an important 
disadvantage of using $\Phi_{p_{T}}$ in the fluctuation analysis. While
$\Phi_{p_{T}}$ is sensitive to the presence of particle correlations 
in a system, it does not provide information on 
the nature of the correlation. Several effects 
can contribute to $\Phi_{p_{T}}$. Therefore, to achieve a better 
understanding of the fluctuation structure, it is useful to also employ a 
more differential method \cite{Tra00}. 

\indent
The correlations can be studied by plotting the cumulative $p_T$ variables
for particle pairs. Namely, instead of $p_{T}$, one can introduce the 
variable $x$, defined for a particle $i$ as \cite{Bia90}

\begin{equation}
x(p_T)=\int_{0}^{p_{T}}\rho({p_{T}}')d{p_{T}}',
\label{def_3}
\end{equation}
where $\rho(p_{T})$ is the inclusive $p_{T}$ distribution, normalized to
unity, which is obtained from all particles used in the analysis. By 
construction, the $x$ variable varies between 0 and 1 with a {\em flat} 
probability distribution. Two-particle correlation plots, as 
presented in this paper, are obtained by plotting $(x_{1},x_{2})$ points 
for all possible particle pairs within the same event. The number of pairs 
in each $(x_{1},x_{2})$ bin is divided by the mean 
number of pairs in a bin (averaged over all $(x_{1},x_{2})$ bins). 
This two-dimensional plot is {\it uniformly} populated when no 
interparticle correlations are present in the system. Nonuniformity 
signals the presence of interparticle correlations. For example, 
when identical particles are studied, Bose statistics lead to a ridge 
along the diagonal of the $(x_{1},x_{2})$ plot, which starts at $(0,0)$ 
and ends at $(1,1)$, whereas event-by-event temperature fluctuations 
produce a saddle-shaped structure \cite{Tra00}.

\section{Experimental Setup}
\label{s:experiment}

\begin{figure}[h]
\begin{center}
\includegraphics[width=6cm, angle=270]{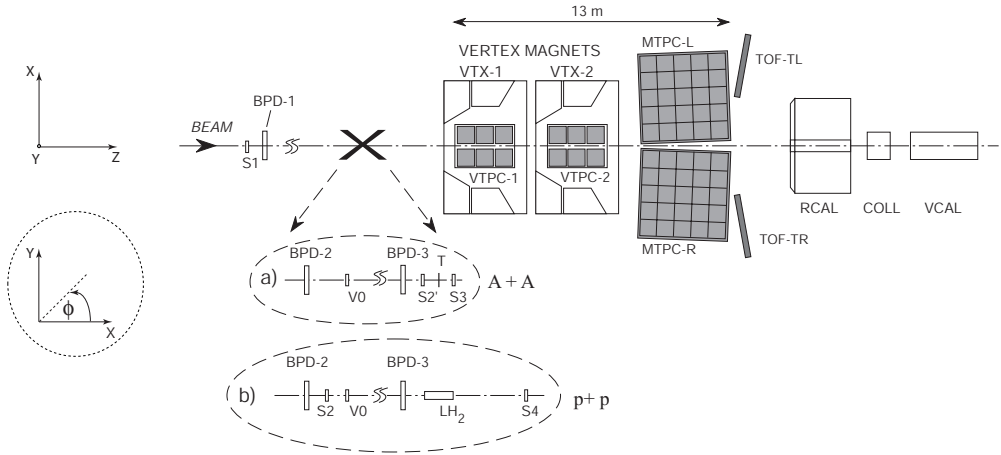}
\end{center}
\vspace{-0.5cm}
\caption{\small Experimental setup of the NA49 experiment
\cite{na49_nim} with different beam definitions and target
arrangements.}                                                           
\label{setup}
\end{figure}

NA49 is one of the fixed target experiments at the CERN SPS. The detector 
(see Fig.~\ref{setup} and Ref. \cite{na49_nim}) is a large acceptance 
hadron spectrometer used for the study of the hadronic final states produced 
in $p+p$, $p+$nucleus, and nucleus+nucleus collisions. In particular, 
centrality selected Pb+Pb interactions were recorded at 20$A$, 30$A$, 
40$A$, 80$A$, and 158$A$ GeV projectile energies.

\indent
The main components of the detector are four large-volume 
time projection chambers (TPCs), which are capable of
detecting about 80\% of approximately 1500 charged particles created in a 
central Pb+Pb collision at 158$A$ GeV (the acceptance losses are concentrated 
mainly in the backward rapidity region and at azimuthal angles along 
the magnetic field direction). 
The vertex TPCs (VTPC-1 and VTPC-2), are located in the
magnetic field of two superconducting dipole magnets (1.5 and 1.1 T, 
respectively, at 158$A$ GeV; for lower energies, the magnetic field 
is scaled down in proportion to the beam energy). Two other TPCs (MTPC-L 
and MTPC-R) are positioned downstream of the magnets symmetrically to the 
beam line. The results presented in this paper are analyzed with a global 
tracking scheme \cite{na49_global}, which combines track segments that 
belong to the same physical particle but were detected in different TPCs. 
The NA49 TPCs allow precise measurements of particle momenta $p$ with a 
resolution of $\sigma(p)/p^2 \cong (0.3-7)\cdot10^{-4}$ (GeV/c)$^{-1}$. 
A precise measurement of specific energy loss ($dE/dx$) in the region of 
relativistic rise is possible in the TPCs, however, $dE/dx$ information 
is not used in this analysis.

\indent 
The centrality of nuclear collisions is selected by using 
the energy of the projectile spectator nucleons measured in the 
downstream calorimeter (VCAL). The geometrical acceptance of
the VCAL is adjusted to cover the projectile
spectator region by setting the collimator (COLL).

\indent
The target is a thin Pb foil (224 mg/cm$^{2}$) positioned about 80 cm 
upstream from VTPC-1. Pb beam particles are identified by means of their
charge as seen by a counter (S2') situated in
front of the target. The beam position detectors (BPD-1/2/3 in Fig. 
\ref{setup}), which are proportional chambers placed along the beam 
line, provide a precise measurement of the transverse positions of the 
incoming beam particles. For Pb beams, interactions in the target are 
selected by an
anti-coincidence of the incoming beam particle with a  
counter (S3) placed directly behind the target.

\indent
Details of the NA49 detector setup and performance of the tracking 
software are described in \cite{na49_nim}.

\section{Data selection and analysis}
\label{s:data}

\subsection{Data sets}

The data used for the analysis consist of samples of Pb+Pb collisions at  
20$A$, 30$A$, 40$A$, 80$A$, and 158$A$ GeV energy for which the 7.2\% 
most central reactions were selected. Table~\ref{data_sets_energy} 
describes the data sets used in this analysis.

\begin{table}[h]
\begin{center}
\begin{tabular}{|c|c|c|c|c|}
\hline
Collision energy (GeV) & Year & No. events & $\sqrt{s_{NN}}$ (GeV) &
$y_{c.m.}$ \cr
\hline
\hline
20 $A$ & 2002 & 229 000 & 6.27 & 1.88 \cr
\hline
30 $A$ & 2002 & 297 000 & 7.62 & 2.08 \cr
\hline
40 $A$ & 1999 & 165 000 & 8.73 & 2.22 \cr
\hline
80 $A$ & 2000 & 228 000 & 12.3 & 2.56 \cr
\hline
158 $A$ & 1996 & 166 000 & 17.3 & 2.91 \cr
\hline
\end{tabular}
\end{center}
\vspace{-0.5cm}
\caption {\small Data sets used in this analysis, for 7.2\% central
Pb+Pb collisions: collision energy, year of data taking, number of
events, center-of-mass energy $\sqrt{s_{NN}}$ for $N+N$ pair,
and center-of-mass rapidity $y_{c.m.}$ in the laboratory frame. }
\label{data_sets_energy}
\end{table}

\subsection{Event and particle selection}
\label{s:selection}

Event selection criteria were aimed at reducing  
possible contamination with nontarget collisions. The primary vertex was
reconstructed by fitting the intersection point of the measured
particle trajectories. Only events with a proper quality and position
of the reconstructed vertex are accepted in this analysis. The 
vertex coordinate $z$ along the beam has to satisfy $|z-z_{0}|<\Delta z$, 
where $z_{0}$ is the nominal vertex position and 
$\Delta z$ is a cut parameter. The values of $z_{0}$ and $\Delta z$ are 
-581.05 and 0.25 cm for 20$A$ GeV, -581.3 and 0.3 cm for 30$A$ GeV, -581.1 
and 0.3 cm for 40$A$ GeV, -581.2 and 0.3 cm for 80$A$ GeV, and -578.9 
and 
0.3 cm for 158$A$ GeV. The maximal allowed deviation of the $x$ and $y$ 
positions of the fitted vertex ($\Delta x$ and $\Delta y$) from 
the measured beam position varies from 0.1 
cm (for 158$A$ GeV) to 0.25 cm (for low energies). The cut on the ratio of 
the number of tracks used to fit the primary vertex divided by the total 
number of tracks registered in TPCs ($ntf/nto$) was required 
to be higher than 0.25 for all energies.

\indent
To reduce the possible contamination by nonvertex tracks such as 
particles from secondary interactions or from weak decays, several track
cuts are applied. The accepted particles are required to have
measured points in at least one of the vertex TPCs. 
A cut on the so-called track impact parameter, the distance between the
reconstructed main vertex and the back extrapolated track in the target 
plane $|b_x|<2$ cm and $|b_y|<1$ cm, is applied to reduce the 
contribution of nonvertex particles. Moreover, particles are accepted 
only when the potential number 
of points ($nmp$), calculated on the basis of the
geometry of the track, in the detector exceeds 30. The ratio 
of the number of points on a track to the potential number of points 
($np/nmp$) is required to be higher than 0.5 to avoid using 
fragments of tracks. This set of 
cuts significantly reduces the possible contamination of particles from 
weak decays, secondary interactions, and other sources.

\indent
In this analysis only tracks with $0.005 < p_T < 
1.5$ GeV/c are used. For all five energies, the forward-rapidity region 
is selected as $1.1 < y^*_{\pi} < 2.6$, where $y^*_{\pi}$ is the particle 
rapidity calculated in the center-of-mass reference system. As the 
track-by-track identification is not always possible in the experiment, 
the rapidities are calculated assuming the pion mass for all particles.

\begin{figure}[h]
\begin{center}
\includegraphics[width=14cm]{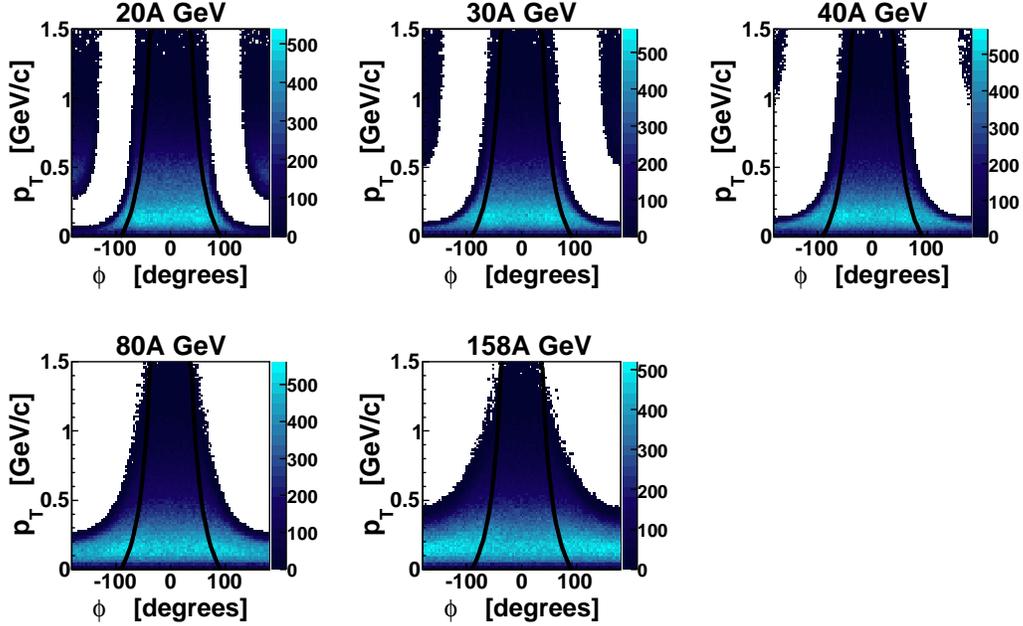}
\end{center}
\vspace{-0.8cm}
\caption {\small (Color online) NA49 ($\phi,p_{T}$) acceptance of all 
charged particles for $2.0 < y_{\pi}^{*} < 2.2$. Additional cut on
$y_{p}^{*}$ (see the text) is not included. The solid lines represent 
the analytical parametrization of the common acceptance.}
\label{azimuth2}
\end{figure}

\indent
The NA49 detector provides a large, though incomplete, acceptance 
in the forward hemisphere. Figure~\ref{azimuth2} presents examples of 
($\phi,p_T$) acceptance \footnote{\label{acc1} All 
charged particles are plotted, with the azimuthal angle of 
negatively charged particles (assuming standard polarity of the magnetic 
field) reflected: namely for all negatively charged particles 
with $\phi<0$ degrees their azimuthal angle is changed as follows: 
$\phi$ goes to $\phi+360$ degrees, and finally for all negatively 
charged particles $\phi$ goes to $\phi-180$ degrees. In the case of the 
opposite polarity of the 
magnetic field (40$A$ GeV data) one has to redefine the azimuthal angle of 
positively charged particles, instead.}  for $2.0 < y^*_{\pi} < 2.2$.
The regions of complete azimuthal acceptance are different for various 
energies. The quantitative comparison of $\Phi_{p_T}$ values 
among the five energies requires selection of the regions of common 
acceptance.

\indent
The solid lines in Fig.~\ref{azimuth2} represent a 
parametrization of the common acceptance limits for all five energies 
by the formula -
\begin{equation}
p_{T}(\phi)=\frac{A}{\phi^{2}}-B,
\label{acc_eq}
\end{equation}
where the parameters $A$ and $B$ depend on the rapidity range as given in 
Table~\ref{a_energy}. The same limits are used for both the negatively 
charged and positively charged particles with the 
appropriate redefinition of azimuthal angle $^{\ref{acc1}}$.
Only particles within the acceptance limits are used in the analysis.

\begin{table}[h]
\begin{center}
\begin{tabular}{|c|c|c|c|c|c|c|c|c|}
\hline
$y_{\pi}^{*}$  & 1.0-1.2 & 1.2-1.4 & 1.4-1.6 & 1.6-1.8 & 1.8-2.0 & 2.0-2.2
& 2.2-2.4 & 2.4-2.6 \cr
\hline
\hline
$A (\frac{deg.^{2}GeV}{c})$ & 600 & 700 & 1000 & 2600 & 3000 & 2500 &
1800 & 1000 \cr
\hline
$B (\frac {GeV}{c})$ & 0.2 & 0.2 & 0.2 & 0.5 & 0.4 & 0.3 & 0.3 & 0.1 \cr
\hline
\end{tabular}
\end{center}
\vspace{-0.5cm}
\caption {\small Parametrization of the NA49 $\phi - p_T$ acceptance
common for all five energies for positively charged particles (standard
configuration of magnetic field). For negatively charged particles one
has to change the definition of the azimuthal angle (see text) and then
use the same parametrization.}
\label{a_energy}
\end{table}

\indent
Due to the described kinematic cuts (mostly on rapidity) and 
track selection criteria, together with the requirement to use the same
limited ($\phi,p_{T}$) acceptance, only about 5.0 - 5.2 \% of 
charged particles produced in central Pb+Pb interactions (at each of the
five energies) enter into the subsequent analysis.

\begin{figure}[h]
\begin{center}
\includegraphics[width=14cm]{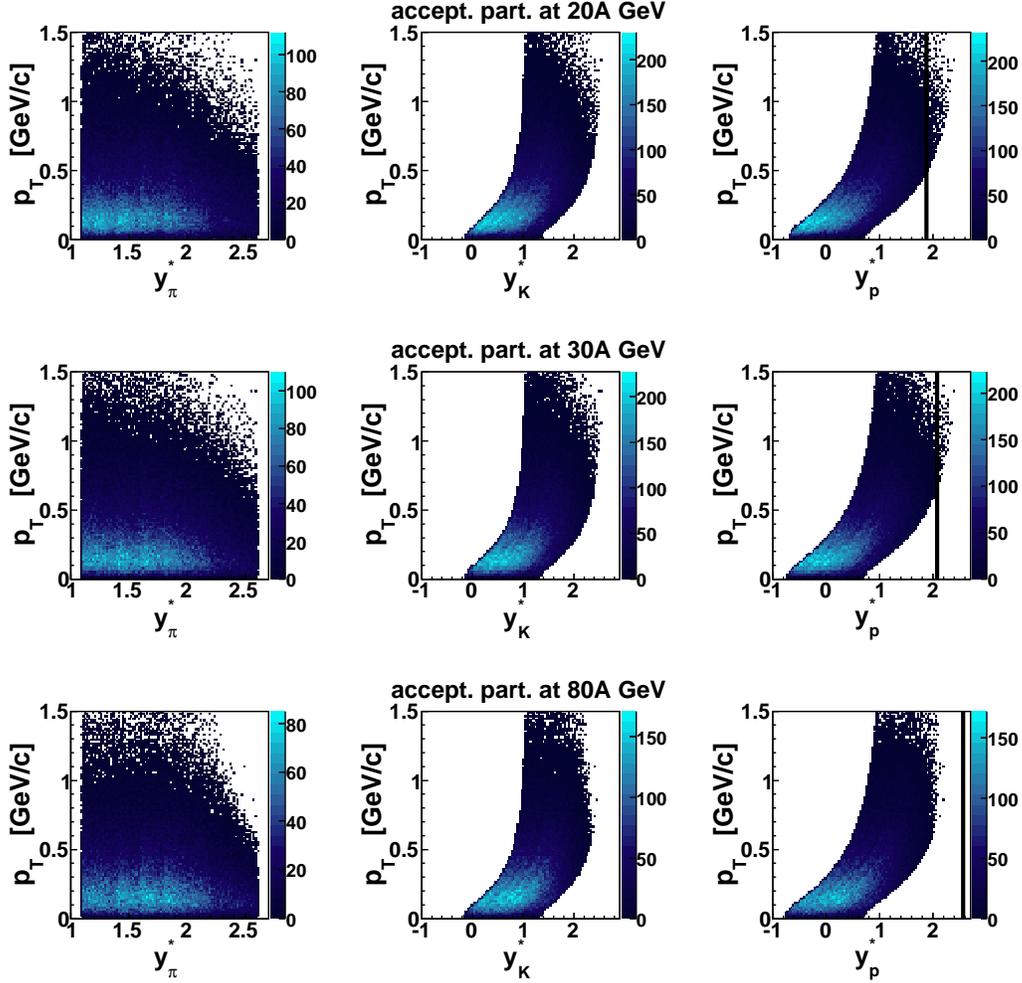}
\end{center}
\vspace{-0.8cm}
\caption {\small (Color online) ($y^{*},p_{T}$) plots of all accepted 
(see Fig. \ref{azimuth2}) particles assuming pion (left), kaon (middle) 
and proton (right) mass.  Additional cut on $y_{p}^{*}$ (see the text) 
is not included. Top, middle, and bottom panels correspond to 20$A$, 
30$A$, and 80$A$ GeV data, respectively. Black lines represent beam 
rapidities ($y^{*}_{beam}$) in the center-of-mass reference system.}
\label{acceptance_ypt}
\end{figure}

\indent
Figure~\ref{acceptance_ypt} presents ($y^{*},p_{T}$) plots of all
charged particles accepted in the analysis (additional cut on 
$y_{p}^{*}$ - see below - not applied in the plots). Top, middle and 
bottom panels correspond to 20$A$, 30$A$, and 80$A$ GeV data, 
respectively, whereas left, middle, and right panels are obtained by 
assuming pion, kaon and proton masses when evaluating rapidities. Black 
lines represent beam rapidities ($y^{*}_{beam}$) in the center-of-mass 
reference system. At lower energies the NA49 TPC acceptance extends
to the projectile rapidity domain and the selected particles may be
contaminated by e.g. elastically scattered or diffractively produced
protons. This domain has been excluded from the analysis by applying an
additional cut on the rapidity $y_{p}^{*}$ of the particles, calculated
with the proton mass. Namely, at each energy, the rapidity $y_{p}^{*}$
is required to be lower than $y^{*}_{beam}-0.5$.

\indent
The above cut was introduced because 
the preliminary analysis (without the additional $y_{p}^{*}$ cut)  
manifested an unexpected effect: the $\Phi_{p_{T}}$ measure 
showed a different behavior for particles of different charges 
\cite{cpod_kg}. $\Phi_{p_{T}}$ was independent
of energy and consistent with zero for negatively charged
particles, but it significantly increased for lower SPS energies
for both all charged and positively charged particles. This effect, 
however, was found to be connected with protons only, and 
can be explained by event-by-event impact parameter fluctuations or,
more precisely, by a correlation between the number of protons 
(nucleons) in the forward hemisphere and the number of protons 
(nucleons) that are closer to the production region 
\cite{impact_urqmd}. For more central events the number of 
forward-rapidity protons is smaller, and consequently, the number of 
protons in the production region is higher. The situation is opposite 
for less central collisions. The existence of those different event 
classes results in the increased $\Phi_{p_{T}}$ values for positively 
charged particles. One can eliminate this trivial source of correlations 
by either centrality restriction or rejection of the beam rapidity 
region \cite{impact_urqmd}. In this analysis, the second method is 
employed by applying a cut on $y_{p}^{*}$ at each energy (see above).

\subsection{Corrections and error estimates}
\label{s:corrections}

\indent
The statistical error on $\Phi_{p_{T}}$ has been estimated as follows. 
The whole event sample was divided into 30 independent subsamples. 
The value of $\Phi_{p_{T}}$ was evaluated for each subsample, and the 
dispersion - $D$ - of the results was then calculated. The 
statistical error of $\Phi_{p_{T}}$ is taken to be equal to $D/\sqrt{30}$.

\indent
The event and track selection criteria reduce the possible systematic
bias of the measured $\Phi_{p_{T}}$ values. To estimate the
remaining systematic uncertainty, the values of cut parameters were varied
within a reasonable range. For a given cut parameter, the "partial" 
systematic error was taken as half of the difference between the 
highest and the lowest $\Phi_{p_{T}}$ value. Two event cuts and two track 
cuts were considered in the analysis. The final estimate of 
the total systematic error on $\Phi_{p_{T}}$ was taken as the 
maximum of the changes resulting from this study.

\begin{figure}[h]
\begin{center}
\vspace{-0.8cm}
\includegraphics[width=8cm]{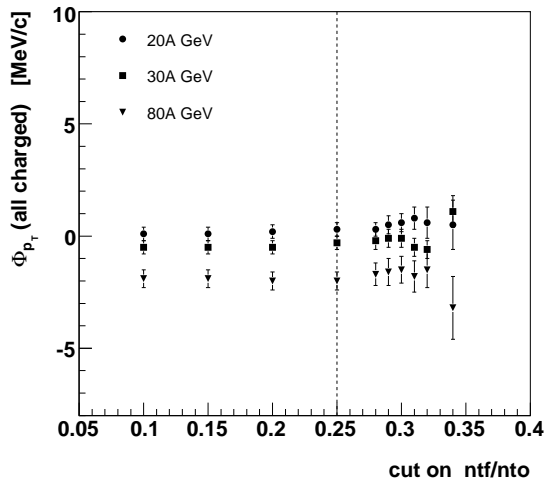}
\end{center}
\vspace{-1.2cm}
\caption {\small Dependence of $\Phi_{p_{T}}$ on one of the event cuts
($ntf/nto$ ratio). Note: the values and their errors are correlated.
The dashed line indicates the cut used in the analysis.}
\label{fipt_ntfnto_energy}
\end{figure}

\begin{figure}[h]
\begin{center}
\vspace{-0.8cm}
\includegraphics[width=8cm]{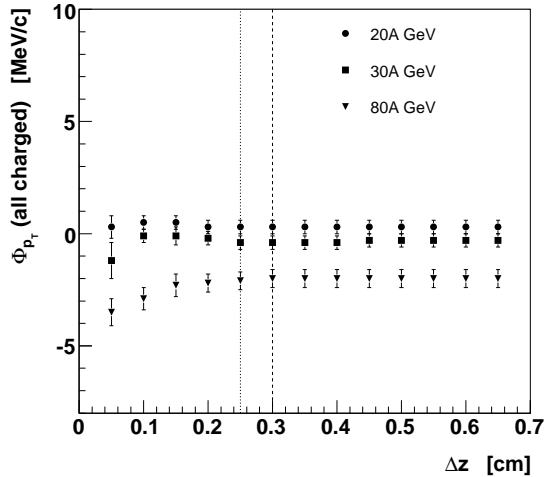}
\end{center}
\vspace{-1.2cm}
\caption {\small Dependence of $\Phi_{p_{T}}$ on the allowed distance
$\Delta z$ from the nominal position of the main vertex. Note: the
values
and their errors are correlated. The dashed line indicates the cut used
in the analysis of 30A and 80A GeV data. The dotted line represents
the cut used in the analysis of 20A GeV data. }
\label{fipt_vz_energy}
\end{figure}

\indent
Event cuts are used to reject possible contamination of
nontarget interactions, however there is always a small fraction of
remaining nontarget events that can influence the $\Phi_{p_{T}}$ values.
This systematic bias can be estimated by investigating the dependence 
of $\Phi_{p_{T}}$ on two vertex cut parameters - $ntf/nto$ and $\Delta 
z$. Figure~\ref{fipt_ntfnto_energy} presents the dependence of 
$\Phi_{p_{T}}$ on the ratio $ntf/nto$ for 20$A$, 30$A$, and 80$A$ GeV data. 
The observed systematic error of $\Phi_{p_{T}}$ with respect to 
changes of $ntf/nto$ varies from 0.35 MeV/c for 20$A$ GeV to 0.85 MeV/c 
for 30$A$ GeV and 80$A$ GeV. In Fig.~\ref{fipt_vz_energy} 
the dependence of $\Phi_{p_{T}}$ on the allowed distance $\Delta z$ from 
the nominal position of the main vertex is shown for three 
energies. The observed variation of $\Phi_{p_{T}}$ with this cut is quite 
small for all studied data sets. The highest "partial" systematic 
error (for 80$A$ GeV data) was found to be 0.75 MeV/c.

\begin{figure}[h]
\begin{center}
\vspace{-0.8cm}
\includegraphics[width=8cm]{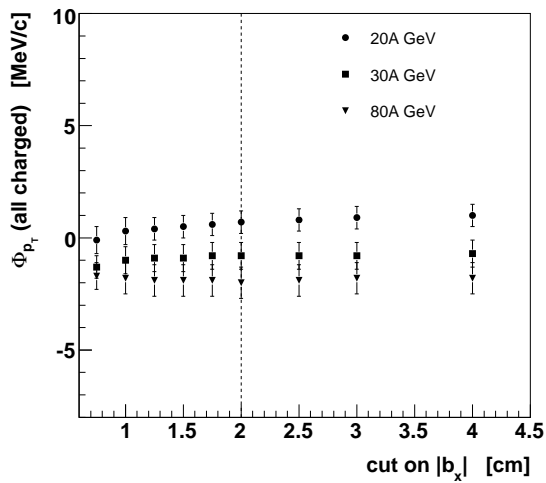}
\end{center}
\vspace{-1.2cm}
\caption {\small Dependence of $\Phi_{p_{T}}$ on upper cut in the
impact parameter $|b_x|$. For each point, the cut on $|b_y|$ is equal
to half the cut on $|b_x|$. Note: the values and their errors are
correlated. The dashed line indicates the cut used in the analysis.}
\label{fipt_bxby_energy}
\end{figure}

\begin{figure}[h]
\begin{center}
\vspace{-0.8cm}
\includegraphics[width=8cm]{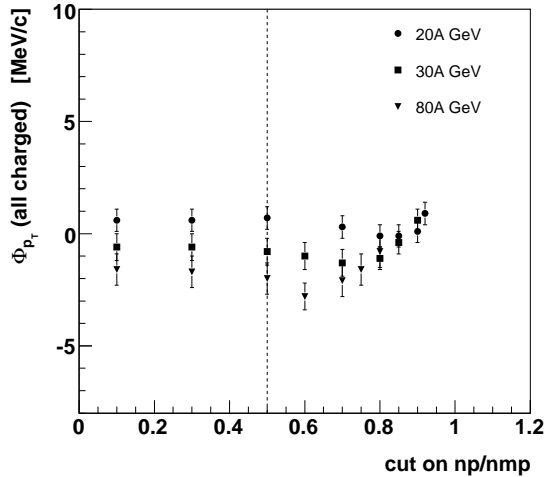}
\end{center}
\vspace{-1.2cm}
\caption {\small Dependence of $\Phi_{p_{T}}$ on lower cut on
$np/nmp$ ratio. Note: the values and their errors are correlated. The
dashed line indicates the cut used in the analysis.}
\label{fipt_npnmp_energy}
\end{figure}

\indent
The majority of tracks selected by the track selection criteria are main
vertex tracks and the remaining fraction ($\approx$10\%) originates 
predominantly from weak decays and secondary interactions with the 
material of the detector. The influence of this remaining fraction can be
estimated by studying the dependence of $\Phi_{p_{T}}$ on the track cut
parameters $|b_x|$ and $|b_y|$ and the $np/nmp$ ratio. Figures
\ref{fipt_bxby_energy} and \ref{fipt_npnmp_energy} present how the
values of $\Phi_{p_{T}}$ change with the impact parameter cut and 
with the cut on the $np/nmp$ ratio, respectively. One finds 
that $\Phi_{p_{T}}$ is rather stable with respect to both cut parameters.
A small increase of $\Phi_{p_{T}}$ with increasing impact parameter cut 
may be due to the increasing contribution of nonvertex tracks for higher 
$|b_x|$ and $|b_y|$ values. This effect has been studied 
quantitatively, for central Pb+Pb collisions at 158$A$ GeV, in our 
previous paper \cite{fluct_size}. The systematic error 
contribution from the impact parameter cut dependence varies from 0.15 
MeV/c for 80$A$ GeV to 0.55 MeV/c for 20$A$ GeV, whereas the 
contribution from the change of cut in the $np/nmp$ ratio varies from 
0.45 MeV/c for 40$A$ GeV and 80$A$ GeV up to 1.45 MeV/c for 158$A$ GeV.

\indent
Finally, the typical systematic error on $\Phi_{p_{T}}$, determined as 
the maximum resulting from the above analysis, has been estimated as not 
higher than 0.7 MeV/c for 20$A$ GeV, 0.9 MeV/c for 80$A$ GeV, 1.1 MeV/c 
for 40$A$ GeV, and 1.6 MeV/c and 1.7 MeV/c for 30$A$ and 158$A$ GeV data, 
respectively. More detailed values (also for various 
charge selections) are given in Table~\ref{data_energy_RAP_CUT}.

\begin{figure}[h]
\begin{center}
\vspace{-0.8cm}
\includegraphics[width=8cm]{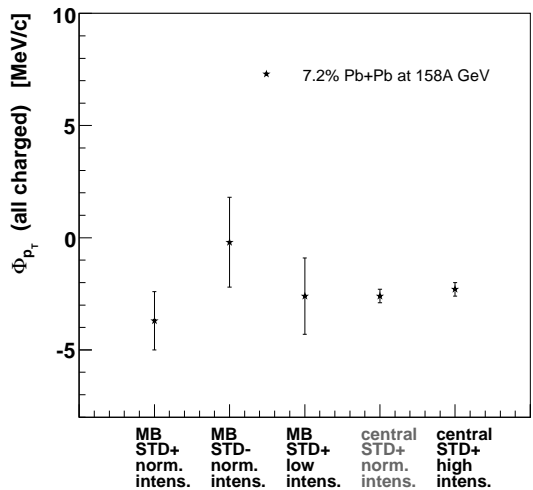}
\end{center}
\vspace{-1.2cm}
\caption {\small $\Phi_{p_{T}}$ values for different central and
minimum-bias (MB) data sets. Data used for the further analysis are
central STD+ normal intensity.}
\label{fipt_prod_type_all}
\end{figure}

\indent
The NA49 experiment registered 158$A$ GeV Pb+Pb interactions at various 
beam intensities, magnetic field configurations ("normal" polarity of 
the magnetic field STD+ and opposite polarity of the magnetic field 
STD-), and with different centrality triggers. 
Figure~\ref{fipt_prod_type_all} presents $\Phi_{p_{T}}$ values for these 
different data sets. An additional cut on the energy deposited in VCAL has been 
applied to select the 7.2\% most central Pb+Pb interactions from each 
data sample. Although the numbers of events extracted from minimum-bias 
data are relatively low, one observes good agreement between the 
$\Phi_{p_{T}}$ values obtained for different experimental conditions.

\begin{figure}[h]
\begin{center}
\vspace{-0.8cm}
\includegraphics[width=8cm]{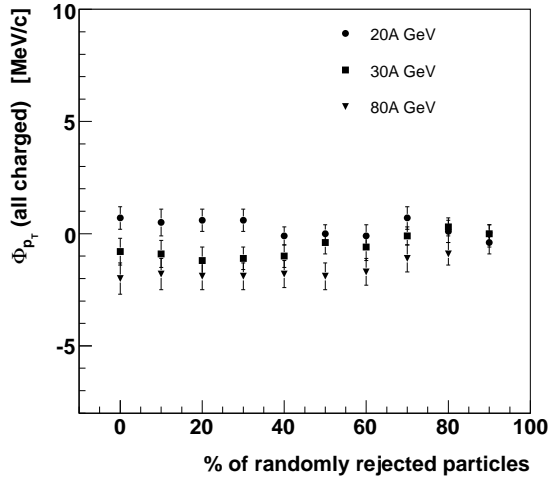}
\end{center}
\vspace{-1.2cm}
\caption {\small Dependence of $\Phi_{p_{T}}$ on the fraction of
randomly rejected particles.}
\label{random_energy}
\end{figure}

\indent
The observed $\Phi_{p_{T}}$ values can be affected by the losses of 
tracks resulting from the reconstruction inefficiency of the 
detector and from the track selection cuts. The dependence of 
$\Phi_{p_{T}}$ on the 
fraction of randomly rejected particles is shown in Fig.~\ref{random_energy}. 
Within the considered forward-rapidity region, at all five energies 
(only three example energies are shown), the tracking efficiency of the 
NA49 detector is higher than 95\% and therefore 
Fig.~\ref{random_energy} implies that the bias due to tracking inefficiency 
and track selection cuts is lower than 0.5 - 1.0 MeV/c.

\indent
It has already been shown in our previous paper \cite{fluct_size} that the
limited two-track resolution influences the measured $\Phi_{p_{T}}$
values. To quantitatively estimate this contribution, five 
samples (30 000 events in each) of mixed events were prepared (for 
20$A$, 30$A$, 40$A$, 80$A$, and 158$A$ GeV data). 
Multiplicities of mixed events were 
chosen to be the same as those of real events, but each particle in a 
mixed event was taken at random from a different real event. Such a 
mixing procedure ensures that the inclusive spectra (e.g., transverse 
momentum or rapidity) are the same for data 
and for mixed events, but all possible correlations vanish. Indeed, it  
was verified for all data sets and all particle charge selections that 
the $\Phi_{p_{T}}$ value calculated for the sample of mixed events is 
consistent with zero. In a second step, the prepared mixed events  
were processed by the NA49 simulation software. The resulting simulated 
raw data were then reconstructed and the $\Phi_{p_{T}}$  measure 
calculated. The obtained $\Phi_{p_{T}}$ values are negative, as expected 
for the anticorrelation introduced by the losses due to the limited 
two-track resolution. The additive two-track resolution correction is 
calculated as the difference ($\Delta\Phi_{p_{T}}$) between the values of 
$\Phi_{p_{T}}$ after detector simulation and reconstruction and before 
this procedure (mixed events). The values of $\Delta \Phi_{p_{T}}$ were 
calculated for all five energies and for various particle charge 
selections, separately, and are listed in Table~\ref{ttr_energy_RAP_CUT}.   

\begin{table}[h]
\begin{center}
\begin{tabular}{|c|c|c|c|c|c|c|}
\hline
Particles  & 20$A$  & 30$A$  & 40$A$  & 80$A$  & 158$A$
\cr
\hline
\hline
All  & -0.6 $\pm$ 0.4  & -0.1 $\pm$ 0.3
& -0.9 $\pm$ 0.4 & -1.7 $\pm$ 0.4 & -3.4 $\pm$ 0.6 \cr
\hline
Negative    & -0.1 $\pm$ 0.2  & 0.4 $\pm$ 0.2 &
-0.7 $\pm$ 0.3 & -1.4 $\pm$ 0.2 & -1.7 $\pm$ 0.5 \cr
\hline
Positive    & -0.2 $\pm$ 0.4 & -0.5 $\pm$ 0.4 &
-0.6 $\pm$ 0.4 & -1.8 $\pm$ 0.5 & -4.2 $\pm$ 0.7 \cr
\hline
\hline
\end{tabular}
\end{center}
\vspace{-0.5cm}
\caption {\small Values of two-track resolution corrections $\Delta
\Phi_{p_{T}}$ (in MeV/c) for all accepted charged particles and
separately for negatively and positively charged particles, for each of
the five energies (in GeV). }
\label{ttr_energy_RAP_CUT}
\end{table}

\indent
The magnitude of the two-track resolution corrections is higher for 
higher energies (from 2 to 4 MeV/c for 158$A$ GeV) where the 
multiplicities of produced particles are higher and consequently the 
densities of tracks are relatively high. The two-track resolution 
corrections become much smaller for low-multiplicity interactions, 
namely for 20$A$ and 30$A$ GeV data ($\Delta\Phi_{p_{T}}$ values close to 
zero). The absolute values of $\Delta \Phi_{p_{T}}$ are typically higher 
for positively than for negatively charged particles, which is mainly 
due to higher track density for positively charged 
particles caused by the larger number of protons relative to antiprotons. The 
$\Delta\Phi_{p_{T}}$ values are typically negative thus indicating that 
$\Phi_{p_{T}}$ measured with an ideal detector would be higher. The 
value of $\Phi_{p_{T}}$ corrected for the limited two-track resolution 
effect equals the "raw" $\Phi_{p_{T}}$ minus the corresponding 
$\Delta\Phi_{p_{T}}$. The error of the corrected $\Phi_{p_{T}}$ is 
calculated by adding in squares the statistical error of the raw 
$\Phi_{p_{T}}$ value and the statistical error of the correction.
As $\Delta\Phi_{p_{T}}$ is very small for 20$A$ and 30$A$ GeV 
data, no corrections for the two-track resolution effect
were applied at these energies. Instead, the upper limits 
of the systematic errors were increased by 0.3 MeV/c (see 
Table~\ref{data_energy_RAP_CUT}).

\section{Results and Discussion}
\label{s:results}

\subsection{Results}

\indent
The results shown in this section refer to {\em accepted} particles, 
i.e., particles that are accepted by the detector and pass all 
kinematic cuts and track selection criteria as discussed in Sec. 
\ref{s:selection}. Results are {\em} not corrected for limited 
kinematic acceptance, and this acceptance has to be taken into account 
when the data are compared with model predictions. The measured 
$\Phi_{p_{T}}$ values are corrected for limited two-track resolution of 
the NA49 detector (see below and Sec. \ref{s:corrections}). A possible 
bias due to particle losses and contamination in the accepted kinematic 
region is included in the systematic error.   

\begin{figure}[h]
\begin{center}
\vspace{-0.5cm}
\includegraphics[width=16cm]{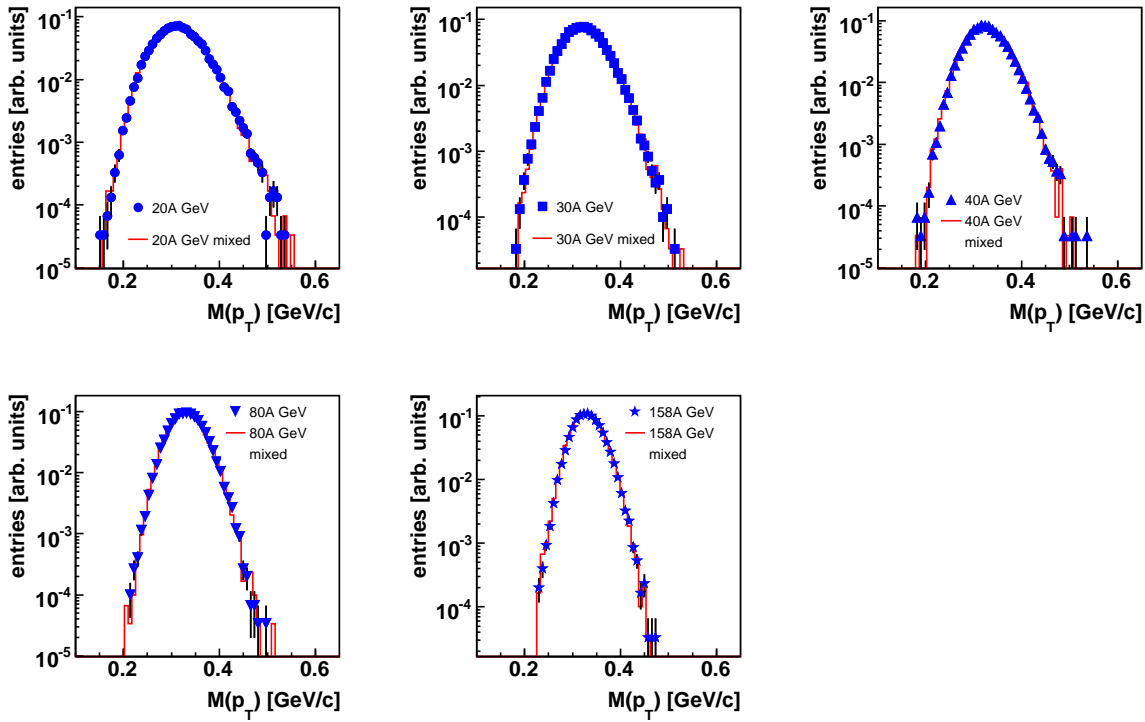}
\end{center}
\vspace{-1cm}
\caption {\small (Color online) Distributions of mean transverse momentum for
data (points) and mixed events (histograms).
Data points are not corrected for acceptance and limited two-track
resolution. Events with accepted particle multiplicity equal to zero are
not used.}
\label{meanpt_energy_RAP_CUT}
\end{figure}

\begin{figure}[h]
\begin{center}
\vspace{-0.5cm}
\includegraphics[width=16cm]{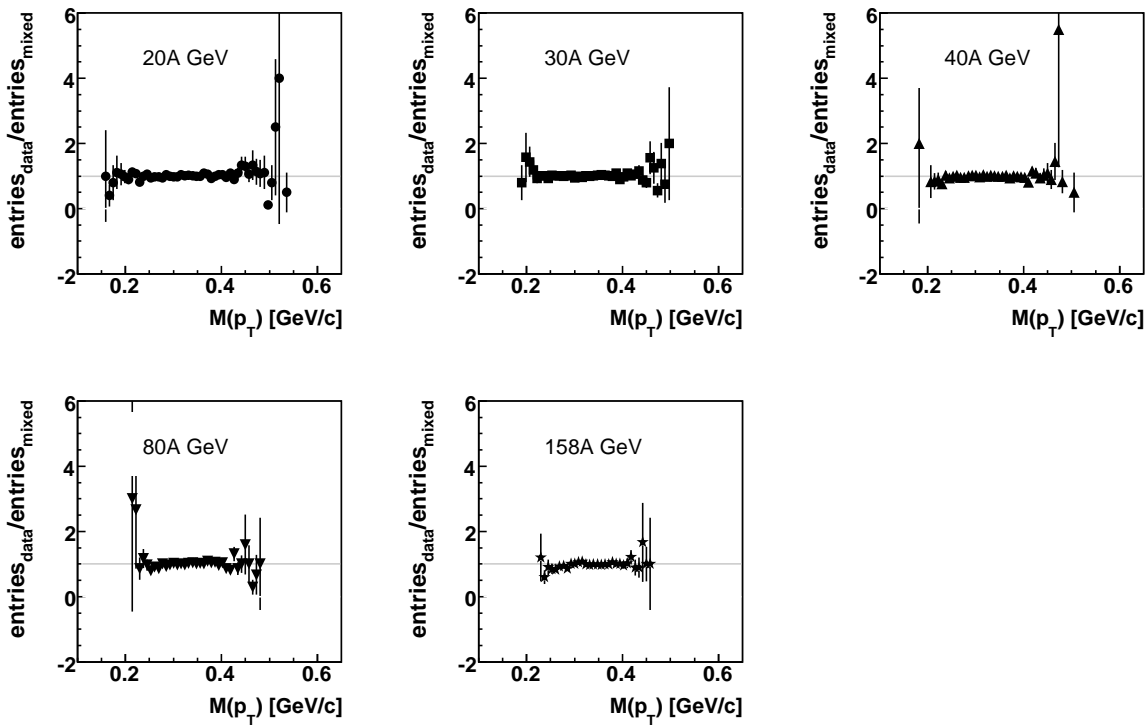}
\end{center}
\vspace{-1cm}
\caption {\small Distributions of mean transverse momentum for data
divided by distribution for mixed events (see the previous figure).}
\label{meanpt_energy_ratio_RAP_CUT}
\end{figure}

\indent
Figure~\ref{meanpt_energy_RAP_CUT} shows the distributions of mean (per 
event) transverse momentum $M(p_T)$ for central 20$A$, 30$A$, 40$A$, 
80$A$ and 158$A$ GeV Pb+Pb interactions. Points represent results from 
real events, and histograms are calculated from mixed events, in which 
there are no interparticle correlations by construction. Data are not 
corrected for experimental effects such as two-track resolution or 
acceptance. Events with zero multiplicity (after cuts, there can
be a few of them especially at lower energies) are not taken into 
account in Figs.~\ref{meanpt_energy_RAP_CUT},
\ref{meanpt_energy_ratio_RAP_CUT}. However, when evaluating 
$\Phi_{p_{T}}$ such events are included in the data sample \footnote 
{It was verified that the $\Phi_{p_{T}}$ value does not depend on whether 
zero multiplicity events are included or not (in agreement with the 
definition of $\Phi_{p_{T}}$). However, in 
Table~\ref{data_energy_RAP_CUT}, the mean multiplicities 
of accepted particles are shown, where events with zero multiplicity 
are normally taken into account. Thus, for consistency, the code 
calculating $\Phi_{p_{T}}$ values includes events with zero multiplicity.}.
All charged particles are used 
to prepare these plots. No significant differences between the 
distributions of $M(p_{T})$ for data and mixed events are observed at 
any SPS energy,  
indicating the absence of substantial event-by-event fluctuations. The 
difference between the histograms for data and mixed events can be 
better seen in Fig.~\ref{meanpt_energy_ratio_RAP_CUT}, which presents 
the ratio of the two. At all energies, the ratio is close 
to unity, thus confirming that average transverse momentum fluctuations are 
very close to the statistical fluctuations. The widths of the 
distributions of $M(p_{T})$ 
decrease with increasing energy of the colliding system as expected 
from the increasing particle multiplicities.

\begin{figure}[h]
\begin{center}
\vspace{-0.8cm}
\includegraphics[width=10cm]{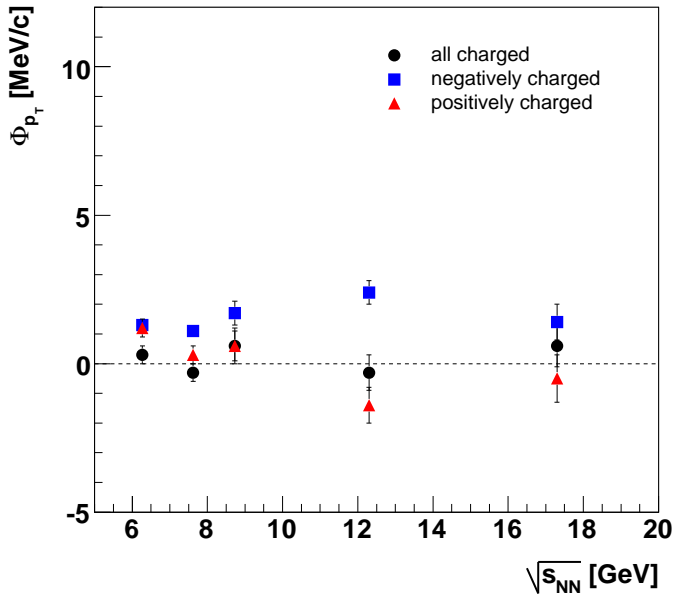}
\end{center}
\vspace{-1.3cm}
\caption {\small (Color online) $\Phi_{p_{T}}$ as a function of energy for the
7.2\% most central Pb+Pb interactions.
Data points are corrected for limited two-track resolution. Errors are
statistical only. Systematic errors are given in Table
\ref{data_energy_RAP_CUT}.}
\label{fipt_energy_RAP_CUT}
\end{figure}

\indent
The fluctuation measure $\Phi_{p_{T}}$, which is more sensitive to small
average $p_T$ fluctuations, is shown in Fig.~\ref{fipt_energy_RAP_CUT}.
The measured values are corrected for limited two-track resolution.
Results are presented for all charged particles and separately for 
negatively and positively charged particles \footnote{The sample of 
negatively charged
particles is composed mainly of negative pions, whereas the sample of
positively charged particles is dominated by positive pions and
protons. Therefore, the measured fluctuations could
differ between both charges. Moreover, the sample of all charged
particles can include additional sources of correlations
which are not present in positively or negatively
charged particles, separately. Therefore the $\Phi_{p_T}$ measure
obtained from all charged particles will not necessarily be
a sum of $\Phi_{p_T}$ for positively and for negatively
charged particles.}. For all three charge selections, $\Phi_{p_{T}}$
seems to be independent of energy and consistent with the hypothesis of
independent particle production ($\Phi_{p_T} \approx 0$). The measured
$\Phi_{p_{T}}$ values do not show any anomalies which might appear when
approaching the phase boundary or the critical point.
However, it should be noted that because of the limited acceptance of 
NA49 and the additional restrictions used for this analysis, the 
sensitivity for fluctuations may be significantly reduced if the 
underlying scale of the correlations is large (see 
Sec.~\ref{s:measures}).

\indent
Table~\ref{data_energy_RAP_CUT} presents the mean multiplicities of
accepted particles, the dispersions $\sigma_N=\sqrt{\langle N^2 \rangle
-\langle N \rangle ^2}$ of the multiplicity distributions, the mean
inclusive transverse momenta, the dispersions $\sigma _{p_T}$ of inclusive
transverse momentum distributions, and $\Phi_{p_{T}}$ values for all
analyzed data sets. The $\Phi_{p_{T}}$ values (with their
statistical and systematic errors) shown in this table have been
calculated for all accepted charged particles as well as for
negatively and positively charged particles, separately. All
$\Phi_{p_{T}}$ values are corrected for the two-track resolution effect
of the NA49 detector.

\clearpage

\begin{table}[h]
\begin{center}
\begin{tabular}{|c|c|c|c|c|c|c|}
\hline
Energy (GeV)  & $ \langle N \rangle $ & $\sigma _N$ & $\overline{p_{T}}$
(MeV/c) & $\sigma _{p_{T}}$ (MeV/c) & $\Phi_{p_{T}}$
({\small \it {$\pm$stat $\pm$sys}})
(MeV/c) \cr
\hline
\hline
20$A$ (all) & 29 & 6 & 314 & 237 & 0.3 $\pm$ 0.3  $^{+0.9}_{-0.6}$ \cr
\hline
20$A$ (-) & 10  & 3 & 221 & 163 & 1.3 $\pm$ 0.2 $^{+0.8}_{-0.5}$ \cr
\hline
20$A$ (+) & 19  & 4 & 361 & 254 & 1.2 $\pm$ 0.3 $^{+1.0}_{-0.7}$ \cr
\hline
\hline

30$A$ (all) & 38 & 6 & 323 & 247 & -0.3 $\pm$ 0.3 $^{+1.3}_{-1.0}$ \cr
\hline
30$A$ (-) & 14  & 4 & 238 & 178 & 1.1 $\pm$ 0.2 $^{+1.0}_{-0.7}$ \cr
\hline
30$A$ (+) & 24  & 5 & 374 & 267 & 0.3 $\pm$ 0.3 $^{+1.9}_{-1.6}$ \cr
\hline
\hline

40$A$ (all) & 44 & 7 & 326 & 251 & 0.6 $\pm$ 0.5 $\pm$ 0.7 \cr
\hline
40$A$ (-) & 17  & 4 & 242 & 182 & 1.7 $\pm$ 0.4 $\pm$ 0.4 \cr
\hline
40$A$ (+) &  27 & 5 & 378 & 273 & 0.6 $\pm$ 0.6 $\pm$ 1.1 \cr
\hline
\hline

80$A$ (all) & 61 & 9 & 331 & 256 & -0.3 $\pm$ 0.6 $\pm$ 0.9 \cr
\hline
80$A$ (-) & 26  & 5 & 270 & 205 & 2.4 $\pm$ 0.4 $\pm$ 0.5 \cr
\hline
80$A$ (+) & 35  & 6 & 378 & 280 & -1.4 $\pm$ 0.6 $\pm$ 0.9 \cr
\hline
\hline

158$A$ (all) & 77 & 10 & 331 & 255 & 0.6 $\pm$ 0.7 $\pm$ 1.5 \cr
\hline
158$A$ (-) & 34  & 6 & 282 & 214 & 1.4 $\pm$ 0.6 $\pm$ 1.0 \cr
\hline
158$A$ (+) & 43  & 7 & 370 & 277 & -0.5 $\pm$ 0.8 $\pm$ 1.7 \cr
\hline
\hline

\end{tabular}
\end{center}
\vspace{-0.5cm}
\caption {\small Measured inclusive and event-by-event parameters for
accepted particles (results with additional $y_{p}^{*}$ cut: see the
text for more details). $\langle N \rangle $, $\sigma _N$,
$\overline{p_{T}}$, and $\sigma _{p_{T}}$ values are not corrected for
acceptance. $\Phi_{p_{T}}$ values are corrected for limited two-track
resolution. $\Phi_{p_{T}}$ values are given with statistical and
systematic errors.}
\label{data_energy_RAP_CUT}
\end{table}

\begin{figure}[h]
\begin{center}
\vspace{-1cm}
\includegraphics[width=14cm]{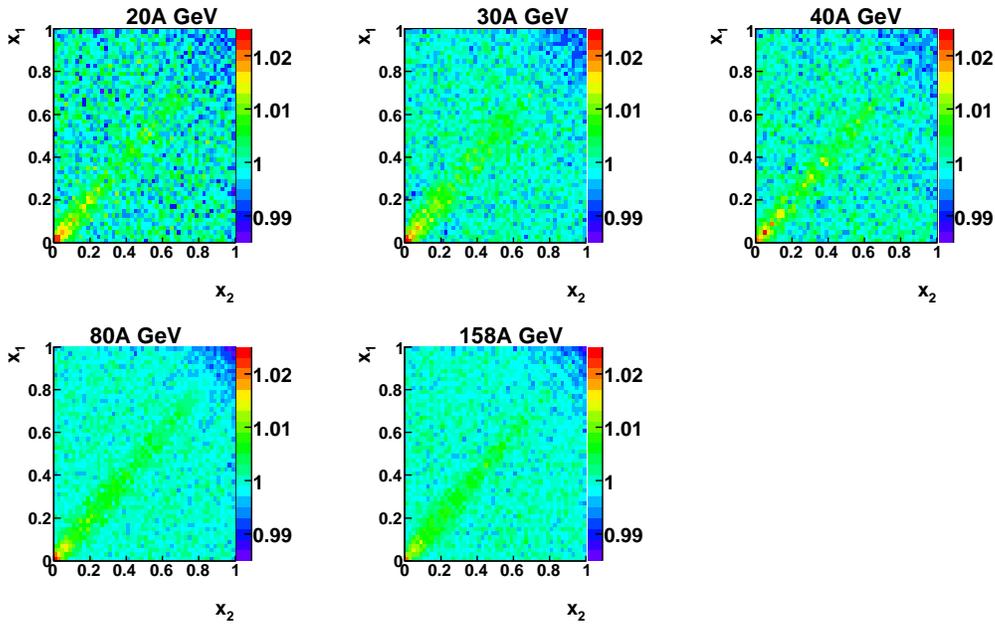}
\end{center}
\vspace{-1cm}
\caption {\small (Color) Two-particle correlation plots $(x_{1},x_{2})$ using
the cumulant $p_T$ variable $x$. The bin contents are
normalized by dividing with the average number of entries per bin.
Plots are for all charged particles for central Pb+Pb
collisions at 20$A$ - 158$A$ GeV.
Note that the color scale is the same in all panels.}
\label{2d_plots_energy_a_RAP_CUT}
\end{figure}

\indent
Two-particle correlation plots (for all charged particles) of the cumulant 
transverse momentum variable $x$ are presented in 
Fig.~\ref{2d_plots_energy_a_RAP_CUT} for 20$A$, 30$A$, 40$A$, 80$A$, and 
158$A$ GeV central Pb+Pb collisions. One
observes that the plots are not uniformly populated. The color scale is 
the same for all plots in order to check whether the correlation pattern 
changes with the energy. For all five energies, short-range correlations 
(Bose-Einstein and Coulomb effects) are visible as an enhancement of the 
point density in the region close to the diagonal. The analysis of 
events simulated by the UrQMD model (see below) resulted in uniformly 
populated two-particle correlation plots without diagonal enhancement.

\begin{figure}[h]
\begin{center}
\vspace{-1cm}
\includegraphics[width=14cm]{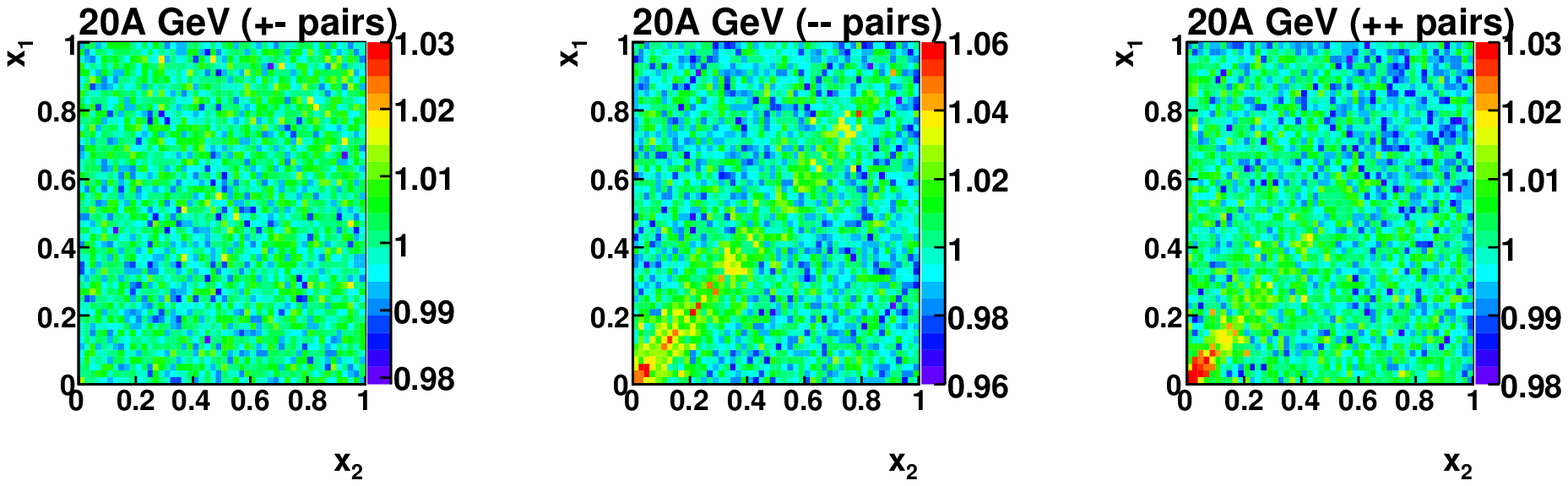}
\end{center}
\vspace{-1cm}
\caption {\small (Color) Two-particle correlation plots $(x_{1},x_{2})$ using
the cumulant $p_T$ variable $x$. The bin contents are
normalized by dividing with the average number of entries per bin.
Plots are for central Pb+Pb collisions at 20$A$ GeV for unlike-sign
particles $(+-)$, and for negatively and positively charged particles,
separately. Note that the color scale can be slightly
different for different panels.}
\label{2D_data_energy_20geV_PM_PP_MM_RAP_CUT}
\end{figure}

\begin{figure}[h]
\begin{center}
\vspace{-1cm}
\includegraphics[width=14cm]{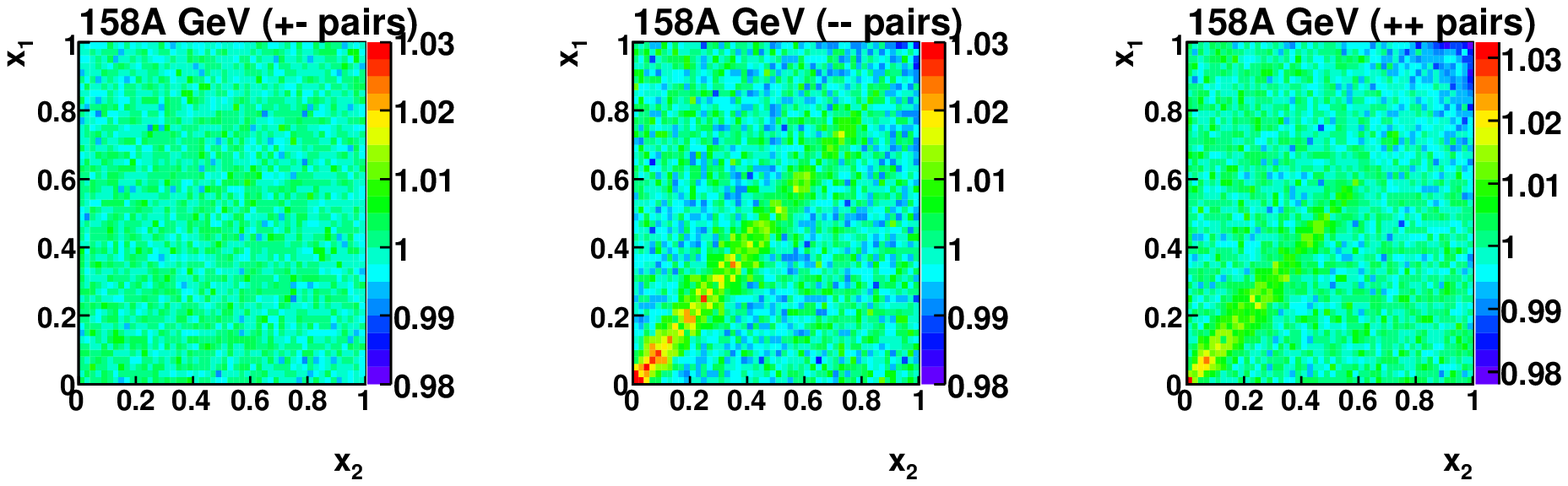}
\end{center}
\vspace{-1cm}
\caption {\small (Color) Two-particle correlation plots $(x_{1},x_{2})$ using
the cumulant $p_T$ variable $x$. The bin contents are
normalized by dividing with the average number of entries per bin.
Plots are for central Pb+Pb collisions at 158$A$ GeV for unlike-sign
particles $(+-)$, and for
negatively and positively charged particles, separately.
Note that the color scale can be slightly
different for different panels.}
\label{2D_data_energy_158geV_PM_PP_MM_RAP_CUT}
\end{figure}

\indent
Figures \ref{2D_data_energy_20geV_PM_PP_MM_RAP_CUT} and 
\ref{2D_data_energy_158geV_PM_PP_MM_RAP_CUT} present two-particle 
correlation plots for different charge combinations. The correlations 
of both positively and negatively charged particle pairs look 
similar (enhancement along diagonal) and can be explained by 
the Bose-Einstein 
effect. On the other hand, there is no significant correlation 
observed for unlike-sign particles. The preliminary analysis for 
$(+-)$ pairs without azimuthal angle restrictions (see the acceptance 
limits in Fig.~\ref{azimuth2}) showed a small maximum restricted to the
region of low $x$ (i.e. low transverse momenta) \cite{phd_kg}. Such a 
low $x$ (low $p_T$) maximum for unlike-sign particles was observed in 
the STAR data \cite{Tra00} and recently also by the CERES experiment 
\cite{ceres_2008}. This maximum was interpreted as an effect of Coulomb 
attraction of particles with different charges and contamination from 
$e^{+}e^{-}$ pairs. When the azimuthal angle is restricted 
(Figs.~\ref{2D_data_energy_20geV_PM_PP_MM_RAP_CUT} and
\ref{2D_data_energy_158geV_PM_PP_MM_RAP_CUT}) the maximum at low $x$ is 
not observed any more.

\clearpage

\subsection{Comparison with the UrQMD model}

\indent
The measured $\Phi_{p_{T}}$ values have been compared with the predictions 
of the ultrarelativistic quantum molecular dynamics
(UrQMD) model \cite{urqmd1, urqmd2}, a transport model producing 
hadrons via formation, decay, and rescattering of resonances 
and strings. In the UrQMD model, no fluctuations due to a phase 
transition are incorporated. However, resonance decays and 
effects of correlated particle production due to energy (momentum) and 
quantum number conservation laws are included. In the analysis default 
parameters of the UrQMD model were used (meson-meson and meson-baryon 
scattering included). For each energy the most central 7.2\% 
interactions were selected, in accordance with the real NA49 events.

\indent
In the study of the NA49 data, the $\Phi_{p_{T}}$ measure was 
calculated from all charged particles, consistent with originating 
from the main vertex. This means that mostly main vertex pions, 
protons, kaons, and their antiparticles are used in the analysis, 
because particles coming from the decays of $K^0_S$, $\Lambda$, $\Xi$ 
and $\Omega$ are suppressed by the track selection cuts. 
Therefore, also in the analysis of the UrQMD events only charged pions, 
protons and kaons were taken into account for evaluating $\Phi_{p_{T}}$.
Due to the specific parameter applied in the UrQMD model - the time 
during which all particles are tracked (taken as 80 fm/c) - the list
of generated kaons, pions and (anti)protons does not contain the
products of weak decays. In the analysis of the UrQMD events, the same 
kinematic restrictions were applied as in the case of the NA49 data. 
In particular, the selected $(\phi, p_T)$  
acceptance is common for all five energies, and as a result of all the 
kinematic and acceptance cuts, only about 5-6\% of all charged particles
are used to study transverse momentum fluctuations.  
 
\begin{figure}[h]
\begin{center}
\includegraphics[width=16cm]{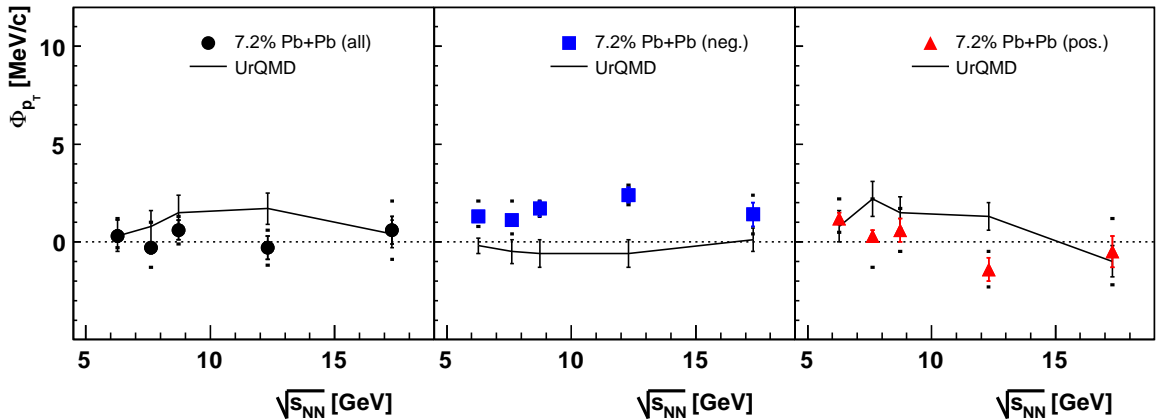}
\end{center}
\vspace{-0.8cm}
\caption {\small (Color online) Comparison of $\Phi_{p_{T}}$ as a function of
energy from data (data points, corrected for limited two-track resolution)
with UrQMD model calculations (black lines) with acceptance
restrictions as for the data.
The panels represent results for all charged (left), negatively
charged (center) and positively charged particles (right). }
\label{fipt_URQMD_onlyone_RAP_CUT}
\end{figure}

\indent
Figure \ref{fipt_URQMD_onlyone_RAP_CUT} compares the the 
energy dependence of $\Phi_{p_{T}}$ for the data and for the UrQMD 
model. Similar to the data, the UrQMD model does not show any 
significant energy dependence of transverse momentum fluctuations for 
any charge selection. The range of $\Phi_{p_{T}}$ values obtained from 
the UrQMD calculations is similar to that found in the data. 
Quantitative comparisons are not conclusive, since in spite of the same 
kinematic and acceptance restrictions the model does not contain the 
effects of Bose-Einstein correlations and Coulomb interactions. 
Correlations implemented in the UrQMD model (due to conservation laws 
and hadron resonance production and decays) apparently do not increase 
the observed $\Phi_{p_{T}}$ values. 
In particular the observed differences between 
model and experimental data do not point to any onset of unusual 
fluctuations as expected for hadronic freeze-out close to the critical 
point.

\vspace{1cm}

\begin{figure}[h]
\begin{center}
\vspace{-0.6cm}
\includegraphics[width=9cm]{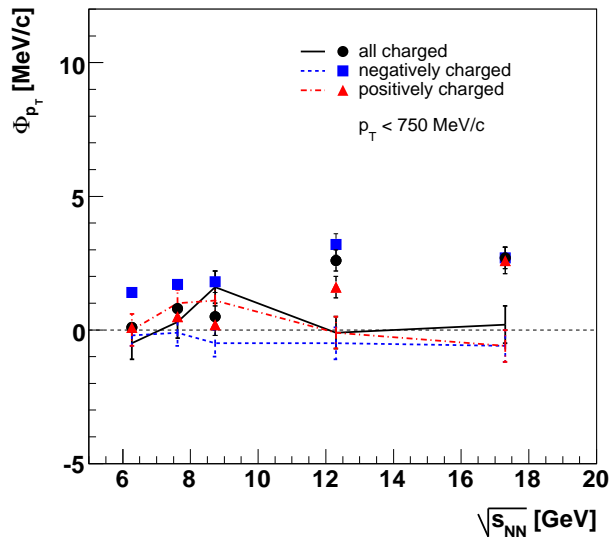}
\end{center}
\vspace{-1.4cm}
\caption {\small (Color online) $\Phi_{p_{T}}$ as a function of energy for the
7.2\% most central Pb+Pb interactions, with additional cut
$p_T < 750$ MeV/c. Data (points) are corrected for limited two-track
resolution. Lines show results of UrQMD calculations with the same
kinematic and acceptance restrictions. Errors are statistical only.}
\label{fipt_energy_RAP_CUT_pt750}
\end{figure}

\begin{figure}[h]
\begin{center}
\vspace{-0.9cm}
\includegraphics[width=9cm]{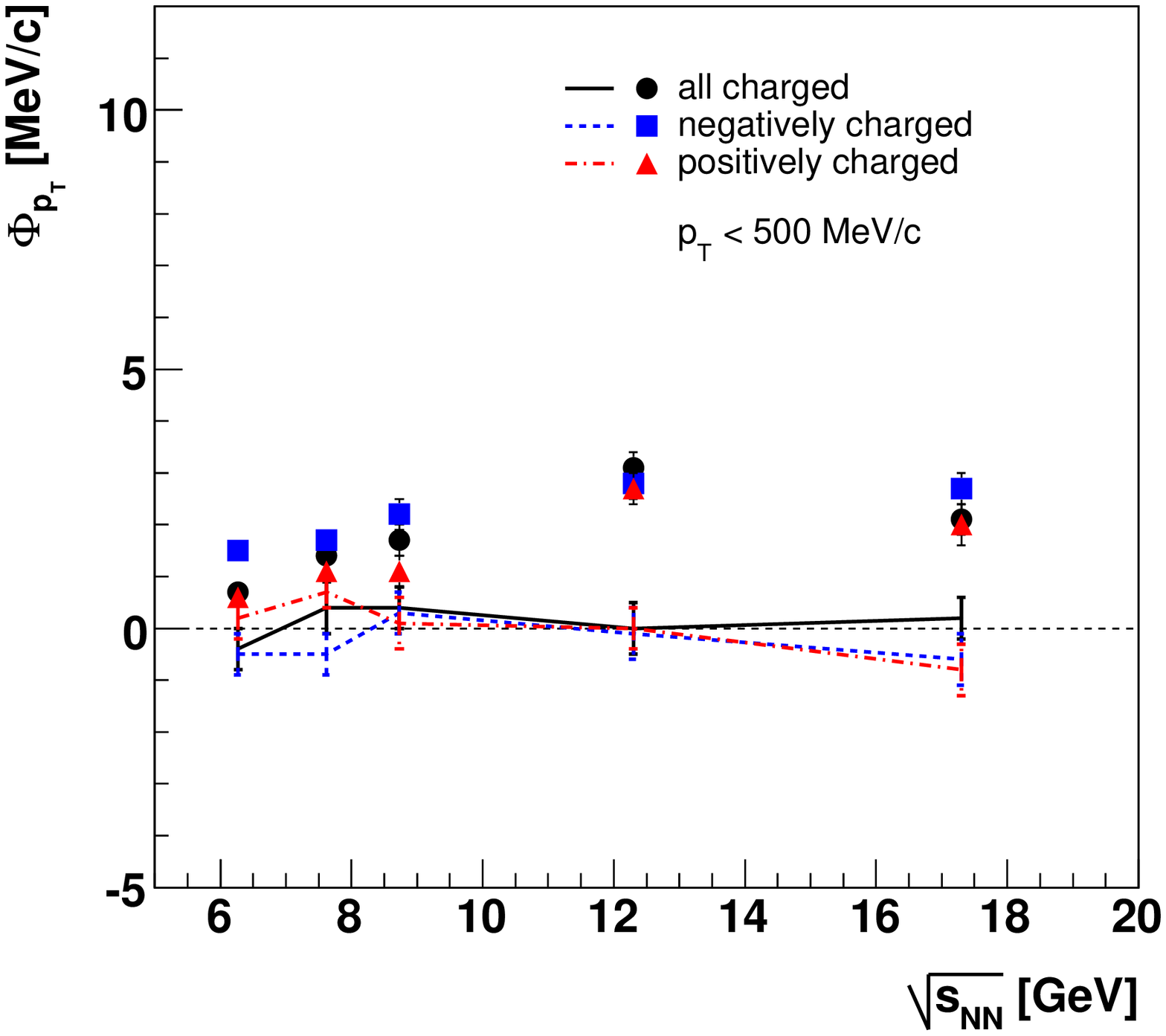}
\end{center}
\vspace{-1.4cm}
\caption {\small (Color online) Same as Fig. \ref{fipt_energy_RAP_CUT_pt750},
but with additional cut $p_T < 500$ MeV/c. }
\label{fipt_energy_RAP_CUT_pt500}
\end{figure}

\begin{figure}[h]
\begin{center}
\vspace{-0.6cm}
\includegraphics[width=9cm]{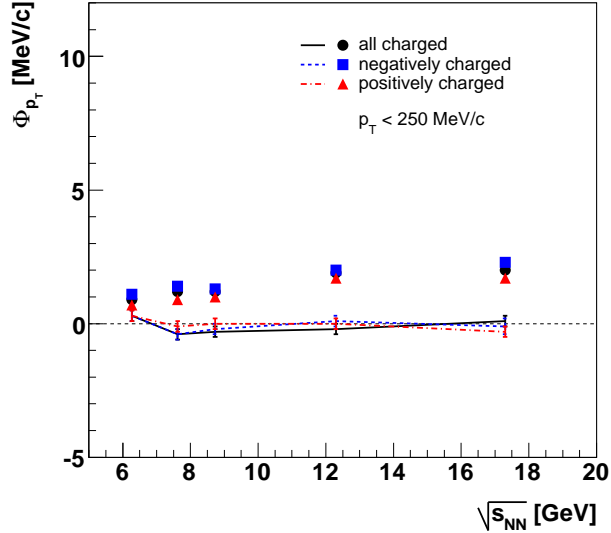}
\end{center}
\vspace{-1.4cm}
\caption {\small (Color online) Same as Fig. \ref{fipt_energy_RAP_CUT_pt750},
but with additional cut $p_T < 250$ MeV/c. }
\label{fipt_energy_RAP_CUT_pt250}
\end{figure}

\indent
The sensitivity of the results on the energy dependence of $\Phi_{p_T}$
to variations of the upper $p_T$ cut was checked by repeating the
analysis for three different upper $p_T$ cuts.
Figures~\ref{fipt_energy_RAP_CUT_pt750}, 
\ref{fipt_energy_RAP_CUT_pt500} and \ref{fipt_energy_RAP_CUT_pt250} show 
the dependence of $\Phi_{p_{T}}$ on energy for upper $p_T$ cuts of 750 MeV/c, 
500 MeV/c and 250 MeV/c, respectively. The measurements (data points) are 
corrected for limited two-track resolution. The lines represent the 
UrQMD predictions with the same kinematic and acceptance restrictions. 
The results with a decreased upper $p_T$ cut do not show any significant 
energy dependence, which is similar to the corresponding results for the 
wide-$p_T$ interval ($p_T < 1500$ MeV/c).
The UrQMD results seem to lie systematically below 
the NA49 data. This effect might be explained by the neglect of 
Bose-Einstein and Coulomb correlations (see also 
Fig.~\ref{2d_plots_energy_a_RAP_CUT}) in the model.

\subsection{Search for the critical point}

\indent
Recent results on the energy dependence of hadron production properties
at the CERN SPS \cite{mg_model, na49_kpi} indicate that deconfined 
matter is formed at the early stage of central Pb+Pb collisions at 
energies as low as 30$A$ GeV. Thus, at higher collision energies, the 
expanding matter crosses the phase boundary between deconfined matter 
and hadron gas and may freeze-out close to it. The nature of the 
transition is expected to change with increasing baryochemical potential 
$\mu_B$. At high potential, the transition is believed to be of the first 
order with the end point of the first order transition line being a 
critical point of the second order. A characteristic property of the 
second-order phase transition is a divergence of the susceptibilities. 
Consequently, an important signal of the critical point are large 
fluctuations, in particular, an enhancement of fluctuations of transverse 
momentum and multiplicity \cite{SRS}. These predictions are to a large
extent qualitative, as QCD at finite temperature and
baryon number is one of the least accessible domains of the theory.
Thus quantitative estimates of the critical point signals were performed
within QCD-inspired models \cite{SRS, crit1, grecy}.

\indent
It was found that assuming a correlation length $\xi = 6$ fm and
freeze-out at the critical point $\Phi_{p_{T}}$ in full phase space may 
be increased by about 40 and 20 MeV/c for all charged and for 
like-charge hadrons, respectively \cite{SRS}. These predictions are 
independent of particle charge and scale with multiplicity. Therefore 
$\Phi_{p_{T}}$ for all charged hadrons is two times larger than for 
like-charge hadrons.
The value of $\Phi_{p_{T}}$ is expected to decrease because of the 
limited acceptance, namely, by a factor of 0.6 and 0.4 
for the NA49 acceptance in rapidity and in azimuthal angle, 
respectively. The particle correlator \cite{SRS} at the critical 
point was calculated to be independent of azimuthal angle and charge of 
hadrons, whereas it changes with rapidity. Therefore the predicted 
$\Phi_{p_{T}}$ values scale with azimuthal acceptance (40\% in this 
analysis, independent of energy (see Fig.~\ref{azimuth2})). Together, 
these reduce the expected signal to about 10 and 5 MeV/c for
all charged and like-charge hadrons, respectively \cite{SRS} (private
communication). These estimates are based on a correlation length $\xi = 
6$ fm. However, because of the finite lifetime of the fireball,
the correlation length may not exceed 3~fm/c \cite{berdnikov}. 
Since the fluctuations scale with the 
square of the correlation length the expected signal is further reduced 
by $(6/3)^{2}=4$. This leads to estimates for $\Phi_{p_{T}}$ of 10 MeV/c 
for all charged hadrons within full phase space and about 2.4 MeV/c for 
all charged hadrons and 1.2 MeV/c for like-sign charged hadrons
in the acceptance of the analysis. Theoretical estimates based on lattice 
QCD calculations locate the critical point at $T \approx$ 162 MeV
and $\mu_B \approx $ 360 MeV \cite{fodor_latt_2004}. Guided by the 
considerations of Ref. \cite{hatta}, we parametrize the increase of 
$\Phi_{p_{T}}$ due to the critical point by Gaussian shapes in $T$ and 
$\mu_B$ with $\sigma(T) \approx$ 10 MeV and $\sigma(\mu_B) \approx$ 30 
MeV, respectively.

\begin{figure}[h]
\begin{center}
\vspace{-0.8cm}
\includegraphics[width=15cm]{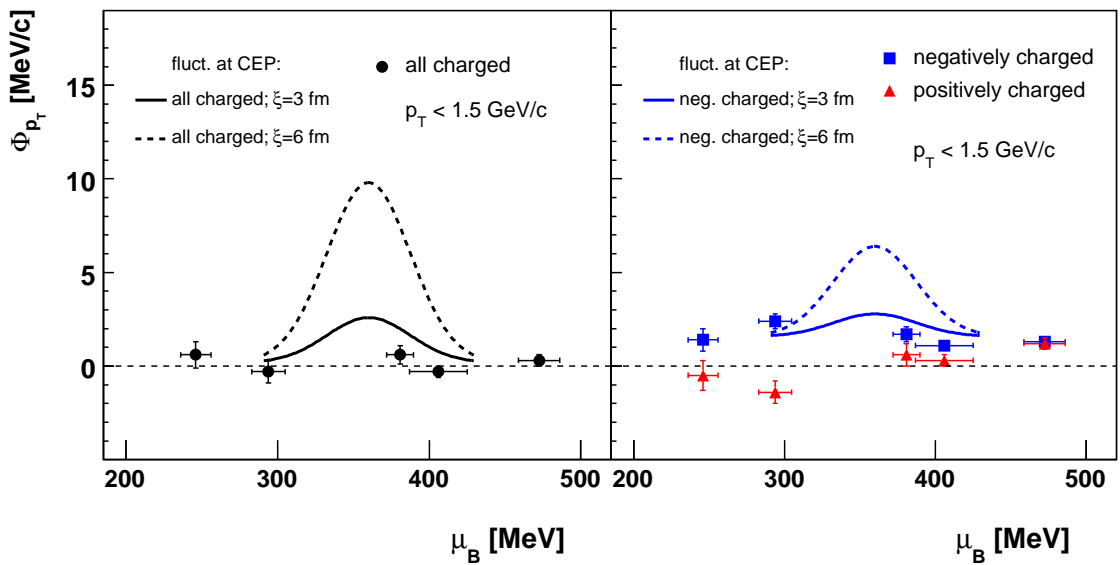}
\end{center}
\vspace{-1.1cm}
\caption {\small (Color online) $\Phi_{p_{T}}$ as a function of baryochemical
potential ($\mu_B$ values from statistical hadron gas model fits
\cite{beccat}) for the 7.2\% most central Pb+Pb interactions. Results
with $p_T < 1.5$ GeV/c (as in Fig.~\ref{fipt_energy_RAP_CUT}). Data
(points) are corrected for limited two-track resolution. Errors are
statistical only. The Gaussian curves show estimated $\Phi_{p_{T}}$
values in the case of the existence of the critical point (see text for
details).}
\label{fipt_miub_RAP_CUT_pt1500_CEPfluct}
\end{figure}

\indent
The results on $\Phi_{p_{T}}$ in central Pb+Pb collisions at the SPS 
energies for $0.005 < p_T < 1.5$GeV/c and the restricted acceptance of
the analysis are shown as a function of $\mu_B$ in 
Fig.~\ref{fipt_miub_RAP_CUT_pt1500_CEPfluct}. The baryochemical 
potential $\mu_B$ at each energy was obtained from fits of the statistical
hadron gas model to the measured particle yields of NA49 \cite{beccat}.
The predictions for the effects of the critical point for the two 
assumptions on the correlation length (3 and 6 fm) are illustrated by 
the dashed and full Gaussian curves, respectively. The base lines of the 
model curves are set to the mean level of $\Phi_{p_{T}}$ in the SPS 
energy range (here the mean level is calculated separately for all 
charged and negatively charged particles). The data do not support the 
predictions for the critical point. Neither a nonmonotonic increase
of $\Phi_{p_{T}}$ nor a characteristic difference between $\Phi_{p_{T}}$
for all charged and like-charge hadrons are observed. However,
the freeze-out in central Pb+Pb collisions appears \cite{beccat} to
take place at temperatures significantly below the transition
temperature and thus could make the critical point signal invisible. 
This possibility motivates the continuation of the search for the 
critical point in NA61/SHINE \cite{na49f_proposal} by a scan in both 
temperature and baryochemical potential which will be performed by 
changing the collision energy and size of the colliding nuclei.

\begin{figure}[h]
\begin{center}
\vspace{-0.8cm}
\includegraphics[width=15cm]{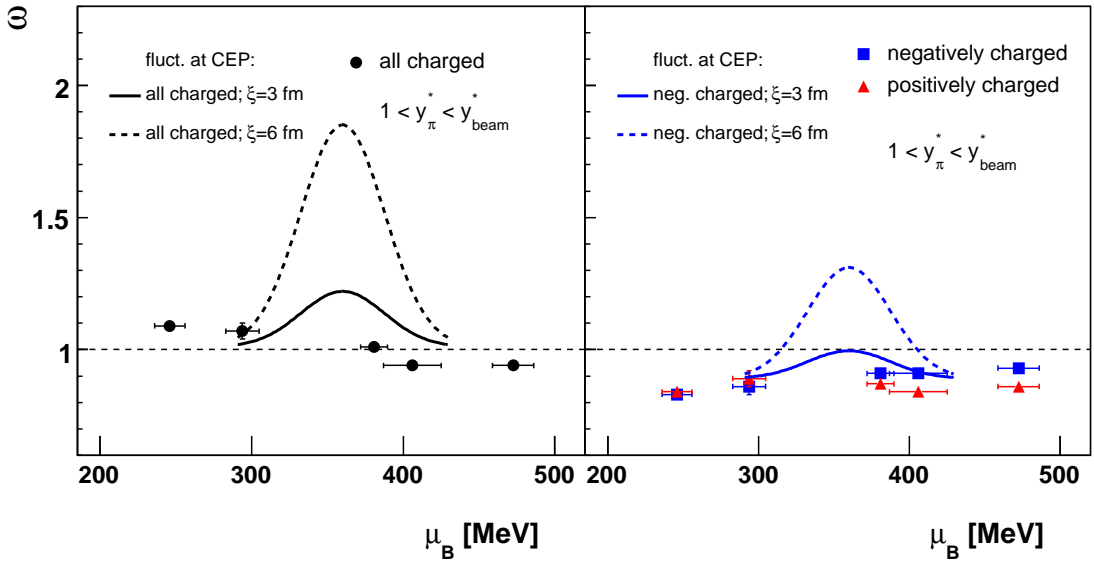}
\end{center}
\vspace{-1.1cm}
\caption {\small (Color online) $\omega$ \cite{benjamin_2007} as a function of
baryochemical potential ($\mu_B$ values from statistical
hadron gas model fits \cite{beccat}) for the 1\% most central Pb+Pb
interactions in $1 < y^{*}_{\pi} < y^{*}_{beam}$. Errors are statistical only.
The Gaussian curves show estimated $\omega$ values in the case of the
existence of the critical point (see text for details).}
\label{omega_miub_CEPfluct}
\end{figure}

\indent
Since publication of the NA49 results on multiplicity fluctuations
\cite{benjamin_2007}, the estimates for the effects of the critical point
have been updated. In the following, we therefore present these
new predictions for multiplicity fluctuations for consistency.
The estimates of the critical point signals were again performed
\cite{SRS} (private communication) assuming first a correlation length 
$\xi = 6$ fm, and they suggest an increase of the scaled variance of the 
multiplicity distribution $\omega$ in full phase space 
by 2 for all charged and by 1 for same-charge particles
(uncorrelated particle production results in $\omega$ = 1).
The limited acceptance used in the multiplicity fluctuation
analysis \cite{benjamin_2007} leads to correction factors for the 
increase of 0.6 and 0.7 due to the rapidity and azimuthal acceptance,
respectively. In the analysis of multiplicity
fluctuations, the azimuthal acceptance was chosen dependent on
the energy with a mean of about 70\%. Thus the measured 
multiplicity fluctuations are expected to be increased by 0.84 and 0.42
for all charged and like-charge hadrons, respectively. For a
correlation length $\xi = 3$ fm, the increase should be four times 
smaller, i.e., 0.5 for all charged and 0.25 for same-charge particles in 
full phase space and, respectively, 0.21 and 0.10 in the NA49 
acceptance.
Figure~\ref{omega_miub_CEPfluct} shows the experimental results for 
$\omega$ as 
a function of baryochemical potential in the rapidity interval $1 < 
y^{*}_{\pi} < y^{*}_{beam}$ for the 1\% most central Pb+Pb collisions
\cite{benjamin_2007}. 
The expectations for the critical point are represented by the Gaussian 
curves above base lines at the level of the mean measured values of 
$\omega$. Similar to transverse momentum fluctuations, multiplicity 
fluctuations do not exhibit the effects expected for hadronic freeze-out 
close to the critical point. However, one should note that the predicted 
width of the critical point signal is comparable to the size of the
$\mu_B$ steps in the NA49 energy scan. In future experiments, one should
consider scanning with narrower steps in $\mu_B$.

\subsection{Comparison with other experiments}

\begin{figure}[h]
\begin{center}
\vspace{-0.8cm}
\includegraphics[width=10cm]{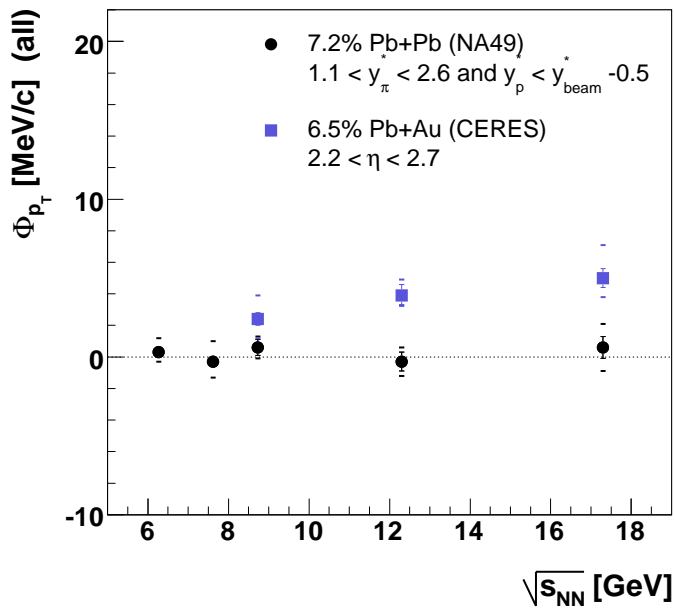}
\end{center}
\vspace{-1.3cm}
\caption {\small (Color online) $\Phi_{p_{T}}$ as a function of energy
measured for all charged particles by the NA49 and CERES experiments.
The NA49 points are obtained for the forward-rapidity region in a limited
azimuthal angle acceptance; the CERES data \cite{CERES} are calculated
for the midrapidity region within a complete azimuthal acceptance. All
data points include small-scale correlations (HBT, etc.).}
\label{comparison_na49_ceres}
\end{figure}

\indent
Event-by-event transverse momentum fluctuations have been studied by
other experiments both at SPS and at RHIC energies. This section 
compares the NA49 results with those from the CERES and the 
STAR experiments. In such comparisons, one has to remember that the 
magnitude of any fluctuation measure may depend on the acceptance used 
in the analysis. The CERES 
experiment measured somewhat higher values of $p_T$ fluctuations in the
midrapidity region for central Pb+Au collisions at 40$A$, 80$A$, and
158$A$ GeV. However, no significant energy dependence was found for
either the $\Sigma_{p_{T}} (\%)$ or the $\Phi_{p_{T}}$ fluctuation 
measures \cite{CERES, ceres_qm2005}. Figure \ref{comparison_na49_ceres} 
compares NA49 and CERES  \cite{CERES} results on the energy dependence 
of the $\Phi_{p_{T}}$ 
measure. One observes only very weak (if any) energy dependence of 
$\Phi_{p_{T}}$ over the whole SPS energy range. It should, however, be 
stressed that quantitative comparison of $\Phi_{p_{T}}$ values from NA49 
and CERES is hampered by different acceptances  
(NA49: forward rapidity and limited azimuthal angle, CERES:
midrapidity and complete azimuthal acceptance).

\begin{figure}[h]
\begin{center}
\vspace{-0.8cm}
\includegraphics[width=10cm]{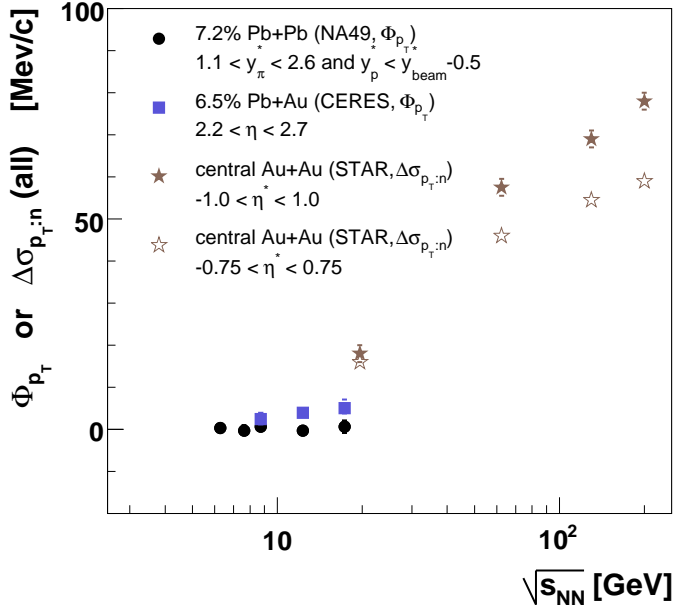}
\end{center}
\vspace{-1.3cm}
\caption {\small (Color online) $\Phi_{p_{T}}$ or $\Delta \sigma_{p_{T}:n}$ as
a function of energy measured for all charged particles by the NA49, CERES,
and STAR experiments. The NA49 points are obtained for the
forward-rapidity region in a limited azimuthal angle acceptance; the CERES
\cite{CERES} and STAR (Fig. 3, left, in Ref. \cite{star_2006b}) points
are calculated for midrapidity regions within complete azimuthal
acceptances. All data points include small-scale correlations (HBT,
etc.). STAR data are corrected for tracking inefficiency.}
\label{comparison_na49_ceres_star}
\end{figure}

\indent
The $\Phi_{p_{T}}$ measure is close to zero at SPS energies, but the 
STAR \cite{star_2006b} results show a strong increase of 
$\Phi_{p_{T}}$ from top SPS to RHIC energies. The energy dependence of 
$\Phi_{p_{T}}$ is shown for central Pb+Pb (Pb+Au, Au+Au) collisions in  
Fig.~\ref{comparison_na49_ceres_star}. The data of NA49 (this paper), 
CERES, \cite{CERES} and STAR \cite{star_2006b} are compiled. The STAR 
results are presented for the related measure $\Delta \sigma_{p_{T}:n}$ 
instead of $\Phi_{p_{T}}$. The difference between both measures 
is smaller than a few percent for high multiplicity collisions as 
considered here (no significant difference within the SPS energy domain and 
only few-percent differences at RHIC energies \cite{star_2006b}).

\begin{figure}[h]
\begin{center}
\vspace{-0.8cm}
\includegraphics[width=10cm]{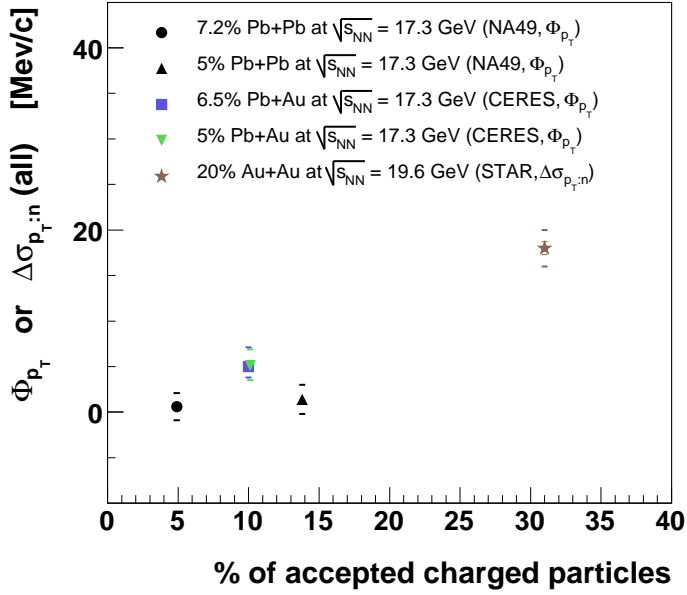}
\end{center}
\vspace{-1.3cm}
\caption {\small (Color online) $\Phi_{p_{T}}$ or $\Delta \sigma_{p_{T}:n}$ as 
a function of percent of charged particles accepted in the analysis. Data
are presented for the top SPS and the lowest RHIC energies.
The NA49 points (Ref. \cite{fluct_size} and this analysis) are obtained for
the forward-rapidity region in a limited azimuthal angle acceptance; the
CERES \cite{CERES} and STAR \cite{star_2006b} points are calculated for
midrapidity regions within complete azimuthal acceptances. All data
points include small-scale correlations (HBT, etc.).}
\label{fipt_fraction_acc_oth_exp}
\end{figure}

\indent
The results of NA49 and CERES at $\sqrt{s_{NN}}$=17.3 GeV 
differ significantly from those of STAR at $\sqrt{s_{NN}}$=19.6 GeV. 
These differences are likely due to differences in the acceptance in 
momentum space. This is illustrated in 
Fig.~\ref{fipt_fraction_acc_oth_exp}, where $\Phi_{p_{T}}$ ($\Delta 
\sigma_{p_{T}:n}$) is plotted as a function of the 
fraction of accepted charged particles~\footnote{The plot shows A+A 
data at the top SPS energy and the lowest RHIC energy. The total 
multiplicities of charged particles ($N_{ch}$) have been obtained from 
the UrQMD model with centralities corresponding to the NA49, the CERES, 
and the STAR data. The following numbers were estimated: $N_{ch}$ = 
1665 for 5\% most central Pb+Pb (NA49), $N_{ch}$ = 1567 for 7.2\% most 
central Pb+Pb (NA49), $N_{ch}$ = 1594 for 5\% most central Pb+Au 
(CERES), $N_{ch}$ = 1575 for 6.5\% most central Pb+Au (CERES), and 
$N_{ch}$ = 1401 for 20\% most central Au+Au (STAR). The values of 
$\Phi_{p_{T}}$ ($\Delta \sigma_{p_{T}:n}$) and multiplicities of 
accepted particles were taken from \cite{fluct_size, CERES, 
star_2006b} and from this paper.}. 
The values of $\Phi_{p_{T}}$ ($\Delta \sigma_{p_{T}:n}$) increase 
with increasing fraction of accepted particles.

\indent
The striking feature of the data presented in 
Fig.~\ref{comparison_na49_ceres_star} is the rapid increase of $\Phi_{p_{T}}$ 
($\Delta \sigma_{p_{T}:n}$) with increasing collision energy. For a 
given experiment, the fraction of accepted charged particles is constant 
(NA49), increases (CERES), or decreases (STAR) with increasing collision 
energy \footnote{The fraction of accepted charged particles in  
NA49 is about 5\%, in CERES it increases from about 
5.8\% at 40$A$ GeV to about 10\% at 80$A$ and 158$A$ GeV, 
and for STAR this fraction decreases from about 31\% at the 
lowest RHIC energy to approximately 17-18\% at the top RHIC energy. In 
order to obtain the above numbers the total multiplicities of charged 
particles were generated within the UrQMD model with centralities 
corresponding to that of the NA49, the CERES, and the STAR data 
presented in Fig.~\ref{comparison_na49_ceres_star}. The multiplicities of {\it 
accepted} charged particles were taken from \cite{CERES}, 
\cite{star_2006b} and from this analysis.}. 
It was already explained in Sec.\ref{s:measures} that 
$\Phi_{p_{T}}$ is independent of acceptance changes provided the 
correlation scale is much smaller than the size of the acceptance 
region. On the other hand, when the range of correlations is 
significantly larger than the acceptance, $\Phi_{p_{T}}$ linearly 
decreases with decreasing fraction of accepted particles. 
The increase of $\Phi_{p_{T}}$ ($\Delta \sigma_{p_{T}:n}$) with energy
is not caused by an increasing magnitude of short range (compared to the 
acceptance) correlations. As mentioned above, these correlations, if
dominant, would lead to an independence of $\Phi_{p_{T}}$ ($\Delta
\sigma_{p_{T}:n}$) of the acceptance, in contrast to the results shown 
in Fig.~\ref{fipt_fraction_acc_oth_exp}. Curiously enough, the increase 
of $\Delta \sigma_{p_{T}:n}$ with growing $\sqrt{s_{NN}}$ for the STAR data 
is associated with a decrease of the fraction of accepted particles. 
Consequently, we conclude that the increase is caused by a {\it growing} 
magnitude of medium- and long-range correlations (correlation length 
comparable to or larger than the acceptance). This effect may result from 
an increasing contribution of particles originating from decays 
of high mass resonances and/or from (mini)jet fragmentation 
\cite{star_2006b}.

\begin{figure}[h]
\begin{center}
\vspace{-0.8cm}
\includegraphics[width=10cm]{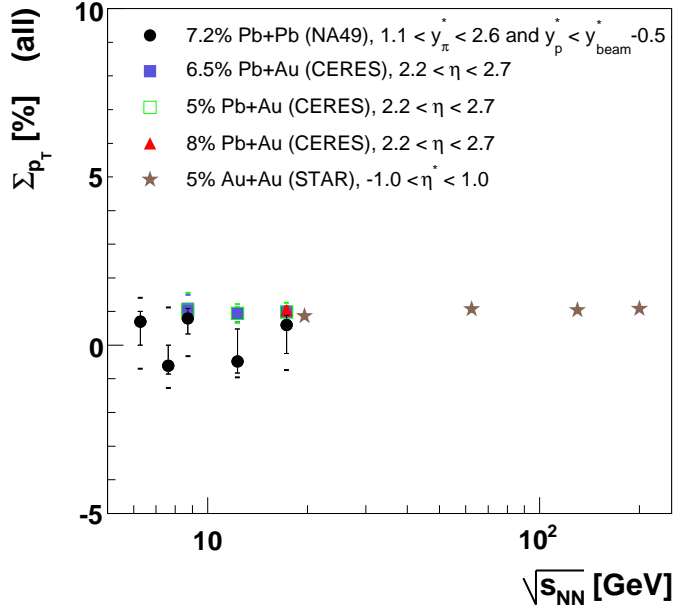}
\end{center}
\vspace{-1.3cm}
\caption {\small (Color online) $\Sigma_{p_{T}} (\%)$ as a function of energy
measured for all charged particles by the NA49, CERES, and STAR experiments.
The NA49 points are obtained for the forward-rapidity region in a
limited azimuthal angle acceptance; the CERES (\cite{CERES}; point for
8\% most central Pb+Au collisions taken from Ref. \cite{ceres_2008}) and
STAR \cite{star_2005} points are calculated for midrapidity regions
within complete azimuthal acceptances. }
\label{sigmapercent_na49_ceres_star}
\end{figure}

\indent
In contrast to the $\Phi_{p_{T}}$ ($\Delta \sigma_{p_{T}:n}$) 
fluctuation measure, $\Sigma_{p_{T}}$ used by the CERES 
experiment, does not show any dramatic differences when going from SPS to 
RHIC energies \cite{CERES, ceres_2008} (see also 
Fig.~\ref{sigmapercent_na49_ceres_star} \footnote{The NA49
$\Sigma_{p_{T}}$ points were estimated from the values of $\Phi_{p_{T}}$ 
using the following equations: $\sigma^2_{p_T, dyn} \cong \frac {2 
\Phi_{p_{T}} \sigma_{p_T}} {\langle N \rangle}$ and finally 
$\Sigma_{p_{T}} \equiv sgn(\sigma^2_{p_T, dyn})
\cdot \frac{\sqrt{|\sigma^2_{p_T, dyn}|}} {\overline{p_T}}$ 
\cite{CERES}. 
For a given $\Delta \Phi_{p_{T}}$ value the errors of $\Sigma_{p_{T}}$ 
(${\Delta \Sigma_{p_{T}}}^{upper}$ and ${\Delta 
\Sigma_{p_{T}}}^{lower}$) were calculated as: ${\Delta 
\Sigma_{p_{T}}}^{upper} = |\Sigma_{p_{T}}(\Phi_{p_{T}} + \Delta 
\Phi_{p_{T}}) - \Sigma_{p_{T}}(\Phi_{p_{T}})|$ and 
${\Delta \Sigma_{p_{T}}}^{lower} = |\Sigma_{p_{T}}(\Phi_{p_{T}} - \Delta
\Phi_{p_{T}}) - \Sigma_{p_{T}}(\Phi_{p_{T}})|$. Finally the NA49 
$\Phi_{p_{T}} \pm \Delta \Phi_{p_{T}}$ corresponds to 
${\Sigma_{p_{T}}}^{+ { \Delta \Sigma_{p_{T}}}^{upper} }_{- {\Delta 
\Sigma_{p_{T}}}^{lower} }$. The same procedure was applied for 
statistical and systematic errors. Although the statistical and 
systematic errors of $\Phi_{p_{T}}$ are comparable at CERES and NA49, 
the errors of $\Sigma_{p_{T}}$ are much higher in NA49. This is due to 
the fact that $\Sigma_{p_{T}}$ is proportional to $sgn(\Phi_{p_{T}}) 
\cdot \sqrt{|\Phi_{p_{T}}|}$ and for $\Phi_{p_{T}}$ close to zero 
(the case of NA49) a small variation of $\Phi_{p_{T}}$ results in much 
higher changes of $\Sigma_{p_{T}}$.  
}). Also the
PHENIX results confirm that the magnitude of the $\Sigma_{p_{T}}
(\%)$ fluctuation measure exhibits only a small variation in a very
wide energy range between $\sqrt{s_{NN}}$ = 22 and 200 GeV
\cite{phenix_qm2006}. It appears that the increase of 
$\Phi_{p_{T}}$ ($\Delta \sigma_{p_{T}:n}$) with the accepted particle 
multiplicity is approximately linear 
(Fig.~\ref{fipt_fraction_acc_oth_exp}). Consequently, $\Sigma_{p_{T}} 
(\%)$, which is proportional to $\sqrt{\Phi_{p_{T}}/ \langle N \rangle}$,  
becomes approximately independent of multiplicity or collision energy.

\indent
A direct comparison of event-by-event $p_T$ fluctuations between 
different experiments, based on the method of determining the magnitude 
of residual temperature fluctuations \cite{Korus2001fv}, has been 
presented in Ref. \cite{mitchel_qm2004}. For the 
most central interactions, the estimated temperature fluctuations are on 
the level of 1.8\%, 1.7\%, 1.3\%, and 0.6\% for PHENIX ($\sqrt{s_{NN}}$ = 200 
GeV), STAR ($\sqrt{s_{NN}}$ = 130 GeV), CERES ($\sqrt{s_{NN}}$ = 17 
GeV), and NA49 ($\sqrt{s_{NN}}$ = 17 GeV), respectively. This 
analysis confirms that, indeed, there is no significant energy 
dependence of temperature fluctuations between SPS and RHIC energies.

\section{Summary and outlook}

\indent
Transverse momentum event-by-event fluctuations were studied
for central Pb+Pb interactions at 20$A$, 30$A$, 40$A$, 80$A$, and 158$A$
GeV. The analysis was limited to the forward-rapidity region
($ 1.1 < y^{*}_{\pi} < 2.6 $). Three fluctuation
observables were studied: the fluctuations of average transverse
momentum ($M(p_T)$) of the event, the $\Phi_{p_{T}}$ fluctuation
measure, and transverse momentum two-particle correlations. The
following results were obtained:
\begin{itemize}
\item
  The distribution of event-by-event average transverse momentum 
$M(p_T)$
  is close to that for mixed events, indicating that fluctuations are
  predominantly of statistical nature.
\item
  The fluctuation measure $\Phi_{p_T}$ has values close to zero and
  shows no significant energy dependence.
\item
  Two-particle transverse momentum correlations are found to be
  negligible except for contributions from Bose-Einstein correlations.
\item
  The UrQMD model reproduces the trend of the data.
\item
  The observed smallness of $p_T$ fluctuations in the SPS energy range
  is consistent with other published measurements.
\item
  No sudden increase or nonmonotonic behavior was observed in the
  energy dependence of $\Phi_{p_T}$ nor the scaled variance $\omega$
  of multiplicity fluctuations. Such effects have been suggested for 
  freeze-out near the critical point of QCD.
\end{itemize}

\indent
Since the limited acceptance of NA49 is expected to 
reduce the signal from possible critical fluctuations, measurements in 
larger acceptance are still needed. Moreover, employing intermediate
size nuclei may possibly move the freeze-out point closer to
the critical point. These considerations motivated several new 
experiments planned in the SPS energy region: NA61 at the CERN SPS
\cite{na49_future, na49_loi, na49f_proposal, newSPSprog}, STAR and
PHENIX at the BNL RHIC \cite{critRHIC_plans}, MPD at the Joint Institute 
for Nuclear Research (JINR) Nuclotron-based Ion Collider Facility (NICA)
\cite{JINR_plans}, and CBM at the GSI Facility for Antiproton and Ion 
Research (FAIR) heavy-ion synchrotron (SIS-300) \cite{FAIR_plans}.
The future results, together with the existing data, will cover a broad
range in the ($T$, $\mu_B$) plane and should lead to significant
progress in the search for the critical point.

\vspace{1cm}
\noindent
{\bf Acknowledgments}

\noindent
This work was supported by
the U.S. Department of Energy Grant DE-FG03-97ER41020/ A000,
the Bundesministerium fur Bildung und Forschung, Germany, 
the Virtual Institute VI-146 of Helmholtz Gemeinschaft, Germany,
the Polish Ministry of Science and Higher Education (1 P03B 006 30, 1 
P03B 127 30, 0297/B/H03/2007/33, N N202 078735),
the Hungarian Scientific Research Foundation (T032648, T032293, 
T043514),
the Hungarian National Science Foundation, OTKA, (F034707),
the Korea Research Foundation (KRF-2007-313-C00175),
the Bulgarian National Science Fund (Ph-09/05),
the Croatian Ministry of Science, Education and Sport (Project 
098-0982887-2878),
and the Stichting FOM, the Netherlands.

\newpage


\end{document}